\documentclass[12pt]{article}
\usepackage{graphicx}
\usepackage{amsmath}
\usepackage{amssymb}
\usepackage{amsbsy}
\usepackage{amsfonts}
\usepackage{setspace}
\usepackage{multirow}
\newcommand{\be}{\begin{equation}}
\newcommand{\ee}{\end{equation}}

\newcommand{\Tr}{\mbox{Tr}}

\usepackage{pdflscape}

\def\a{\alpha}
\def\b{\beta}

\def\d{\delta}

\def\e{\epsilon}
\def\l{\lambda}

\def\p{\partial}
\def\t{\theta}

\def\N{{\cal N}}
\def\O{{\cal O}}

\def\I{{\cal I}}
\def\wt{\widetilde}

\newcommand{\fund}[1]{
\setlength{\unitlength}{#1mm}
\begin{picture}(5,5)(0,.5)
\put(0,5){\line(1,0){5}}
\put(5,5){\line(0,-1){5}}
\put(5,0){\line(-1,0){5}}
\put(0,0){\line(0,1){5}}
\end{picture}}

\newcommand{\twoanti}[1]{{\hskip.1em}
\setlength{\unitlength}{#1mm}
\begin{picture}(6,8)(0,2.5)
\put(0,0){\line(1,0){5}}
\put(0,5){\line(1,0){5}}
\put(0,10){\line(1,0){5}}
\put(0,0){\line(0,1){10}}
\put(5,0){\line(0,1){10}}
\end{picture}}

\newcommand{\twoantic}[1]{{\hskip.1em}
\setlength{\unitlength}{#1mm}
\begin{picture}(6,8)(0,2.5)
\put(0,0){\line(1,0){5}}
\put(0,5){\line(1,0){5}}
\put(0,10){\line(1,0){5}}
\put(0,0){\line(0,1){10}}
\put(5,0){\line(0,1){10}}
\put(-0.2,12){\line(1,0){5.2}}
\end{picture}}

\newcommand{\twosym}[1]{{\hskip.1em}
\setlength{\unitlength}{#1mm}
\begin{picture}(8,6)(2,0)
\put(0,0){\line(0,1){5}}
\put(5,0){\line(0,1){5}}
\put(10,0){\line(0,1){5}}
\put(0,0){\line(1,0){10}}
\put(0,5){\line(1,0){10}}
\end{picture}}

\begin{document}

\baselineskip 0.7cm
\numberwithin{equation}{section}
\begin{titlepage}

\begin{flushright}
IPMU11-0208
\end{flushright}

\vskip 1.35cm
\begin{center}
{\large \bf
The R\"omelsberger Index, Berkooz Deconfinement,\\
and Infinite Families of Seiberg Duals
}
\vskip 1.2cm
Matthew Sudano\vskip 0.2cm

{\it  Institute for the Physics and Mathematics of the Universe, \\
University of Tokyo, Chiba 277-8583, Japan}

{\it School of Natural Sciences, Institute for Advanced Study, \\
Princeton, NJ 08540, USA}

\vskip 1.5cm

\abstract{
R\"omelsberger's index has been argued to be an RG-invariant and, therefore, Seiberg-duality-invariant object that counts protected operators in the IR SCFT of an $\N=1$ theory.  These claims have so far passed all tests. In fact, it remains possible that this index is a perfect discriminant of duality.  
The investigation presented here bolsters such optimism.  
It is shown that the conditions of total ellipticity, which are needed for the mathematical manifestation of duality, are equivalent to the conditions ensuring non-anomalous gauge and flavor symmetries and the matching of (most) 't Hooft anomalies.  Further insights are gained from an analysis of recent results by Craig, et al.  It is shown that a non-perturbative resolution of an apparent mismatch of global symmetries is automatically accounted for in the index.  It is then shown that through an intricate series of dynamical steps, the index not only remains fixed, but the only integral relation needed is the one that gives the ``primitive'' Seiberg dualities, perhaps hinting that the symmetry at the core is fundamental rather than incidental.  
}
\end{center}
\end{titlepage}

\setcounter{page}{2}

\section{Introduction}

Supersymmetric theories were observed to have extraordinary renormalization properties \cite{Wess:1973kz,Wess:1974jb,Iliopoulos:1974zv,Grisaru:1979wc} soon after their discovery.  What is remarkable, however, is not the form of their dynamical evolution but their relative tractability.  The exact results that are accessible in supersymmetric theories have repeatedly substantiated long-held beliefs about the behavior of strongly coupled non-supersymmetric theories.  In this sense, though supersymmetry is widely believed to play an important role at high energies, the field has already yielded dividends.

One of the key advances in opening this window onto the strong dynamics of field theories is Seiberg duality, which relates distinct but IR-equivalent theories.  I think it is fair to say that there is neither a general first-principles derivation of this duality nor a rigorous algorithm for identifying dual theories.  String theory can claim some success in these endeavors \cite{Hanany:1996ie, Elitzur:1997fh, Schmaltz:1997sq, Csaki:1997qe}.  There have also been efforts in the purely four-dimensional setting at systematization.  For example, in \cite{Pouliot:1998yv} Pouliot made an early effort along these lines by constructing a duality-invariant spectroscope of sorts that contained information about the chiral ring of a theory\footnote{The literature on counting-functions of this sort appears to be too vast to cite responsibly.  To the interested reader, I offer only a few of the more modern and seemingly relevant ones \cite{Benvenuti:2006qr, Feng:2007ur, Chen:2011wn} as a starting point.}.  

In this work we will discuss a refinement of this program.  In particular, we will consider an augmented Witten index \cite{Witten:1982df} for a radially quantized superconformal theory on $S^3\times \mathbb{R}$ \cite{Sen:1985ph,Sen:1989bg,Festuccia:2011ws} as defined in \cite{Romelsberger:2005eg,Kinney:2005ej,Romelsberger:2007ec}.  In \cite{Kinney:2005ej} the index was constructed for general $\N$ and was shown to count only protected operators.  The index was also shown to be maximally refined, giving all information that can be determined from group theory\footnote{Implicit in this statement is that all protected information about symmetries outside of the superconformal algebra should also be accounted for.  See \cite{Zwiebel:2011wa} for a proposed further refinement accounting for charge conjugation.}.  

R\"omelsberger discussed only $\N=1$ theories.  He offered a prescription that followed from a free-field analysis, but then permitted the $R$-symmetry to be arbitrary.  It is that object that I refer to as the R\"omelsberger index.  To be clear, if we were to compute the index for the free theory, in which a chiral multiplet has $R[\Phi]=2/3$ and our BPS condition is satisfied for the contributing components, we would get a different answer from one that employs any other $R$-symmetry.  This is as expected because the perturbation that takes us away from the free theory will not in general respect the superconformal symmetry (in particular, the part that we have used to define our BPS condition), and so it does not leave the augmented Witten index invariant.  Reservations have been expressed about the status of this index as a superconformal index \cite{Dolan:2008qi} and even about the claim that it is RG-invariant \cite{anon}.  It would be very satisfying to resolve these confusions, but that is not the topic of this work\footnote{See \cite{Gadde:2010en} and references therein for some discussion of regularization and renormalization in such theories.}.  

For now we take heart from the plethora of experimental evidence \cite{Romelsberger:2007ec, Dolan:2008qi,Spiridonov:2008zr,Spiridonov:2009za,Spiridonov:2011hf,Gadde:2010en} supporting the claim that this object is identical when computed in different theories with a common IR fixed point.  In a perturbative expansion it was shown in \cite{Romelsberger:2007ec} that the indices computed in s-confining $N_c=2$, $N_f=3$ SQCD agree.  Dolan and Osborn then provided the bridge between the physics and math literature and demonstrated {\it exact} equality for the dual pairs of SQCD \cite{Seiberg:1994pq,Intriligator:1995ne} (and equality of SQCD with an adjoint \cite{Kutasov:1995ve, Kutasov:1995np, Kutasov:1995ss} in a large-$N$ limit). The R\"omelsberger index was shown to be given by multidimensional complex integrals of the elliptic gamma functions as introduced by Spiridonov \cite{spiriOne, spiriTwo}. Amazingly, the precise integral identities needed were available and had been proven \cite{2003math9252R} only a few years earlier.  The coincidence appears to be due to the imposition of {\it total ellipticity} \cite{2010arXiv1003.4491S}, which is related to a particular sort of modular invariance, appears to be required for non-trivial integral transformations, and has been conjectured \cite{Spiridonov:2009za} to be a necessary and sufficient condition for 't Hooft anomaly matching \cite{'tHooft:1980xb}.  In fact, the index equality associated with $N_c=2$ $N_f=3$ SQCD was first found in the math literature \cite{spiriOne} as the simplest totally elliptic system.

One of the results of this paper is the demonstration that total ellipticity is indeed related to 't Hooft anomalies.  In fact, it also demands that there are no gauge anomalies (as mentioned in \cite{Vartanov:2010xj}), and that the flavor symmetries are non-anomalous.  We will see, however, that the current definition (as I have understood it) fails to enforce the pure-$R$ 't Hooft anomalies, $\Tr U(1)_R$ and $\Tr U(1)_R^3$.

Other symmetries of these integrals are also interesting.  A particularly famous example generates a 72-member family of theories that includes $N_c=2$, $N_f=4$ SQCD.  In a nice paper by Khmelnitsky \cite{Khmelnitsky:2009vc} it was shown that this family includes the known theories, theories that are obtained from the known ones by perturbing by a relevant operator, and physically inequivalent theories that are ultimately obtained through field redefinitions\footnote{The reason that field redefinitions look different in this case has to do with the presence of accidental symmetries.  There is an ambiguity in the matching of operators \cite{Khmelnitsky:2009vc}.}.  Nevertheless, it is encouraging that (1) none of these theories is actually wrong and (2) purely mathematical considerations correctly (re)produced Seiberg dual theories.  There remain many other provocative claims for new dualities \cite{Spiridonov:2009za,Spiridonov:2010hh,Spiridonov:2011hf} that await scrutiny.

In this work, we will not be so ambitious.  Instead we pursue the goal of systematization following in the footsteps of certain early duality hunters.  In particular, we consider the elliptic hypergeometric integral realization of Berkooz deconfinement \cite{Berkooz:1995km}.  The trick is to interpret a two-index anti-symmetric tensor as a meson, replacing it with fundamentals in a confining gauge theory.  Duality then follows from the basic set \cite{Seiberg:1994pq,Intriligator:1995id,Intriligator:1995ne}.  This method was extended in \cite{Luty:1996cg} to all two-index tensors and reformulated in terms of all-purpose ``modules'' with which a given two-index tensor is replaced.  The deconfined theories, which have product gauge groups, can have infinite families of duals by alternately dualizing the factors of the gauge group. It has been proposed \cite{Spiridonov:2009za} that this phenomenon is given by some sort of Bailey chain \cite{2004TMP139536S} for the associated integrals.  Such theories are among the few excluded from the catalogue of Spiridonov and Vartanov \cite{Spiridonov:2009za,Spiridonov:2011hf}.

\noindent {\it Outline and Summary of Results}

Section \ref{romsec} is primarily devoted to reviewing the construction, establishing notation, and describing the physical significance of the mathematical manipulations that we will need.  This section includes my interpretation of the condition of total ellipticity \eqref{ggg}-\eqref{g}, which is discussed further in appendix \ref{ellapp}.  In section \ref{antisec} the index equality associated with the two-index anti-symmetric deconfinement module of \cite{Luty:1996cg} is demonstrated.  This particular equality is remarkable in several ways:
\begin{enumerate}
\item There is a free positive integer $K$ that determines the rank of the UV gauge group.  The IR theory is ungauged and independent of $K$.
\item Classically, there is an $SU(K)$ symmetry in the UV theory that is absent in the IR.  Upon performing the integral, the parameters associated with this symmetry automatically cancel out.
\item The non-anomalous $R$-symmetry is undetermined.  The index equality holds for any $R$-symmetry, not just the superconformal $R$-symmetry.
\item The required integral identity descends from one of the basic set \cite{2003math9252R} that were also used by Dolan and Osborn in \cite{Dolan:2008qi}.
\end{enumerate}
We go on to apply this result in section \ref{ceht} to the recent work \cite{Craig:2011tx}\footnote{This analysis was generalized in \cite{Craig:2011wj}.}, where we again find that all index equalities follow from those in \cite{2003math9252R}.  This is also true for the $SU(N)$ adjoint deconfinement module \cite{Luty:1996cg}, which is demonstrated in appendix \ref{adjapp}.

\section{The R\"omelsberger Index and Elliptic \\Hypergeometric Integrals}\label{romsec}

Among the formulas presented in this section are the generalized conditions of total ellipticity.  These have not appeared before in the literature; though they have surely been computed \cite{Spiridonov:2009za}.  Nothing else is new.  An introduction to the index and the related mathematics is given.  There have already been several good introductions, including those in \cite{Dolan:2008qi,Spiridonov:2009za,Gadde:2010en}, so I will try not to linger too long on the basics.  Useful formulas are deposited here to establish notation and provide a convenient reference.

\subsection{The R\"omelsberger Prescription}

The index of interest is an augmented Witten index \cite{Witten:1982df} of the form \cite{Kinney:2005ej,Romelsberger:2005eg,Romelsberger:2007ec}
\be\label{indexdef}
I(p,q,y,z)=\Tr(-1)^Fe^{-\b{\cal H}}p^{R/2+J_3+\bar J_3}q^{R/2-J_3+\bar J_3}f(y)g(z).
\ee
Let's now unpack this definition.  Of course, the first operator, $(-1)^F$, is the familiar fermion number operator.  Our index is to be computed for a radially quantized supersymmetric field theory with $R$-symmetry $R$ on a three-sphere \cite{Sen:1985ph,Sen:1989bg,Festuccia:2011ws} for which $J_3$ and $\bar J_3$ are the Cartan generators of the $SU(2)\times SU(2)$ isometry group of the $S^3$.  The combinations of these operators that appear are chosen to commute with ${\cal H}$.  Only states annihilated by ${\cal H}$ contribute to the index.  It is chosen to be of the form
\be
{\cal H}=\{\cal Q,\cal Q^+\},
\ee
where ${\cal Q}$ is one of the supercharges, say ${\cal Q}=\bar Q_1$, and the adjoint is ${\cal Q}^+=-\bar S_1$.  These are the superconformal supercharges $Q$ and $S$ that respectively anti-commute to give translations and special conformal transformations,
\be
\{Q,\bar Q\}\sim P,\qquad \{S,\bar S\}\sim K.
\ee
The full algebra can be found in many places, including, for example \cite{Kinney:2005ej}. Our operator ${\cal H}$ can be re-expressed as
\be\label{H}
{\cal H}={H-2\bar J_3-\frac32R},
\ee
where $H$ is the dilatation operator.

The final operators in \eqref{indexdef} are generators of other commuting non-anomalous symmetries.  $f(y)$ and $g(z)$ respectively are elements of the flavor and gauge groups.  They are given explicitly as $\exp(\a^aT_c^a)$, where the $T_c^a$ are Cartan generators, and so $a$ runs over the rank of the group.  The relationship between $\a$ and $y$ (or $z$) is discussed in appendix \ref{grouptheoryapp}, where we deposit other useful bits of group theory.  We perform a standard change of variables so that the trace (the characters of the representation) are conveniently normalized polynomials.  For example, the fundamental of $U(3)$ is taken to have character $\chi_{U(3),\Box}=u_1+u_2+u_3$. 
This is nearly all one needs to know to evaluate \eqref{indexdef}.  The trace is, or course, a trace over the Hilbert space, and the projection onto gauge invariant states is left implicit.  

Let's now be more explicit. We start by identifying the states that contribute.  We are instructed in \cite{Romelsberger:2007ec} to begin by considering the fields of our UV theory as free fields.  For free fields, looking back at \eqref{H} we find that the scalar component of a chiral multiplet, which has scaling dimension 1, spin 0, and $R$-charge 2/3, is annihilated by ${\cal H}$ (and hence ${\cal Q}$ and $\cal Q^+$).  The other type of matter field that contributes is a right-chiral free fermion, which has quantum numbers $(H=3/2, \bar J_3=1/2, R=1/3)$.  Finally, derivatives with $\bar J_3=1/2$ contribute.  

Now allowing for an arbitrary non-anomalous $R$-charge, $r$, as per the prescription, the single-letter index for a matter superfield transforming in the $\l_{F(G)}$ representation of the flavor (gauge) group is 
\be\label{letter}
i_\Phi(p,q,y,z)=\frac{(pq)^{r/2}\chi_{F,\l_F}(y)\chi_{G,\l_G}(z)-(pq)^{(2-r)/2}\chi_{F,\bar\l_F}(y)\chi_{G,\bar\l_G}(z)}{(1-p)(1-q),}
\ee
where the denominators account for arbitrary derivatives of type $\p_{++}$ and $\p_{-+}$, and $\chi_{H,\l}$ is the character of the representation $\l$ of the group $H$.

After a similar analysis and a bit of simplification the gauge-multiplet single-letter index is found to be
\be
i_V(p,q,z)=-\bigg(\frac{p}{1-p}+\frac q{1-q}\bigg)\chi_{G,Adj}(z).
\ee
All compositions of these letters into words are then given by the plethystic exponential, 
\be
PE\left[f(u_1,u_2,\dots)\right]=\exp\bigg(\sum_{n=1}^\infty\frac1n f(u_1^n,u_2^n,\dots)\bigg).
\ee
Finally, we restrict to gauge singlets by integrating over the gauge group.  These details will be discussed further in the next subsection.  

We are told to interpret this index as the superconformal index of the IR theory.  As suggested in \cite{Gadde:2010en} one should make sure to select the superconformal $R$-symmetry for this to make sense.  If necessary, a-maximization \cite{Intriligator:2003jj} may be used.  Evidence for interpreting this index as a superconformal index has come from observing an expansion of the form
\be\label{scinterpret}
I(t,x,y,z)=n_0+\sum_{a}n_a\frac{t^{q_a}\chi_{SU(2),a}(x)\chi_{F,a}(y)}{(1-tx)(1-tx^{-1})},\qquad p=tx,\qquad q=tx^{-1}
\ee
for integers $n_a$, giving the spectrum of protected operators.  See \cite{Dolan:2008qi} for further discussion of this point, including the allowed forms of the parameters.

\subsection{Calculating the Index}

The recipe of the last section can be summarized in a line.  The index is defined for any supersymmetric four-dimensional field theory as
\be
I(p,q,y)=\int_Gd\mu(z)\exp\left\{\sum_{n=1}^\infty\frac1n\left[i_V(p^n,q^n,z^n)+\sum_\Phi i_\Phi(p^n,q^n,y^n,z^n)\right]\right\}.
\ee 
The only data needed to calculate this index for a given theory is contained in the ubiquitous and much-loved tables such as \eqref{twoantitable}.  With the transformation properties of the fields specified, one just has to look up the characters and the group measure.  Formulas for the characters and Haar measures that we will need are collected in appendix \ref{grouptheoryapp}.  

If one is content to evaluate several orders in an expansion in $t^2=pq$, this completes the story.  However, it was recently shown \cite{Dolan:2008qi} that the full index can be written in terms of special functions known as elliptic gamma functions, and recent results in the math literature \cite{2003math9252R} allowed them to prove exact equalities of indices for Seiberg dual theories. 

The most common representation of the elliptic gamma functions is in terms of a double infinite product. This arises from evaluating the sum in the exponential in terms of logs and then expressing the result in terms of a product of more exotic special functions.  This procedure is straightforward.  It requires only the power series of log and its derivative,
\begin{align}
\ln(1-u)&=-\sum_{n=1}^\infty \frac1nu^n,\\
\frac1{1-u}&=\sum_{n=0}^\infty u^n.\label{geo}
\end{align}
For example, for a generic matter field, we unpack the denominators using \eqref{geo}: 
\begin{align}
i_\Phi(p^n,q^n,u^n)
&=\frac{(pq)^{nr/2}\chi(u^n)-(pq)^{n(2-r)/2}\chi(u^{-n})}{(1-p^n)(1-q^n)}\\
&=\sum_{a,b\ge0}\left[(pq)^{nr/2}\chi(u^n)p^{na}q^{nb}-(pq)^{n(2-r)/2}\chi(u^{-n})p^{na}q^{nb}\right].
\end{align}
This reduces to a finite sum of infinite geometric series upon expanding the character in terms of monomials,
\be
\chi(u)=\sum_c\eta_c(u),\qquad\chi(u^n)=\sum_c\eta_c(u)^n.
\ee
We then find sums of logs in our exponential.
\begin{align}
\I_\Phi(p,q,u)
&\equiv PE\big[i_\Phi(p^n,q^n,u^n)\big]\nonumber\\
&=\exp\left\{\sum_{a,b\ge0}\sum_c\sum_{n=1}^\infty\frac1n\left(\left[(pq)^{r/2}\eta_c(u)p^{a}q^{b}\right]^n-\left[(pq)^{(2-r)/2}\eta_c(u)^{-1}p^{a}q^{b}\right]^n\right)\right\}\nonumber\\
&=\prod_c\prod_{a,b\ge0}\frac{1-[(pq)^{r/2}\eta_c(u)]^{-1}p^{1+a}q^{1+b}}{1-(pq)^{r/2}\eta_c(u)p^{a}q^{b}}.
\end{align}
The double infinite product is a special function known as an elliptic gamma function.  In fact, we will need the following family of special functions.
\begin{align}
\Gamma(u;p,q)&=\prod_{a,b\ge0}\frac{1-u^{-1}p^{a+1}q^{b+1}}{1-up^aq^b},\label{gamma}\\
\theta(u;p)&=\prod_{a\ge0}(1-up^a)(1-u^{-1}p^{a+1}),\\
(u;p)&=\prod_{a\ge0}(1-up^a).
\end{align}
For convenience we will employ a few other standard shorthands that are common in the literature.
\begin{align}
\Gamma(u)&=\Gamma(u;p,q)\\
\Gamma(u_1,u_2,\dots,u_n)&=\prod_{a=1}^n\Gamma(u_a)\\
\Gamma(u_1u_2^{\pm s})&=\Gamma(u_1u_2^s,u_1u_2^{-s})\\
\Gamma(u_1u_2^{\pm s}u_3^{\pm s})&=\Gamma(u_1u_2^su_3^s,u_1u_2^su_3^{-s},u_1u_2^{-s}u_3^s,u_1u_2^{-s}u_3^{-s})
\end{align}
We will discuss the properties of these functions further in the next subsection.  For now we just observe that the contribution to the integrand of the index from a single matter field---a single-field word, if you like---may be written simply as
\be\label{word}
\I_\Phi(p,q,u)=\prod_a\Gamma\left((pq)^{r/2}\eta_a(u)\right).
\ee
In practice then the $R$-charge is inserted along with the character monomials, and the sum that defined the character is replaced by a product over the elliptic gamma functions.  For example, a quark superfield transforming as a fundamental under the gauge group $SU(N_c)$ and as an anti-fundamental under a flavor $SU(N_f)$ contributes
\be
\I_Q(p,q,u)=\prod_{a=1}^{N_f}\prod_{b=1}^{N_c}\Gamma\left((pq)^{r_Q/2}y_a^{-1}z_b\right),\qquad \prod_{a=1}^{N_f} y_a=\prod_{a=1}^{N_c}z_a=1\,.
\ee

The vector field contributions are slightly more tedious to calculate, but they need only be worked out once for each group.  Since they do not change, and since they simplify when combined with the Haar measure it is useful to define {\it gauge theory measures} as follows.
\begin{align}
[dz]_{SU(N)}&\equiv\frac{1}{N!}\left(\prod_{a=1}^{N-1}\frac{dz_a}{2\pi iz_a}{(p;p)(q;q)}\right)\frac1{\prod_{1\le b<c\le N}\Gamma(z_bz_c^{-1},z_b^{-1}z_c)}\bigg|_{\prod_{a=1}^Nz_a=1}\nonumber\\
[dz]_{Sp(2N)}&\equiv\frac{1}{N!}\left(\prod_{a=1}^N\frac{dz_a}{4\pi iz_a}\frac{(p;p)(q;q)}{\Gamma(z_a^2,z_a^{-2})}\right)\frac1{\prod_{1\le b<c\le N}\Gamma(z_bz_c,z_bz_c^{-1},z_b^{-1}z_c,z_b^{-1}z_c^{-1})}\nonumber\\
[dz]_{SO(2N)}&\equiv\frac2{N!}\left(\prod_{a=1}^N\frac{dz_a}{4\pi iz_a}(p;p)(q;q)\right)\frac1{\prod_{1\le b<c\le N}\Gamma(z_bz_c,z_bz_c^{-1},z_b^{-1}z_c,z_b^{-1}z_c^{-1})}\nonumber\\
[dz]_{SO(2N+1)}&\equiv\frac1{N!}\left(\prod_{a=1}^N\frac{dz_a}{4\pi iz_a}\frac{(p;p)(q;q)}{\Gamma(z_a,z_a^{-1})}\right)\frac1{\prod_{1\le b<c\le N}\Gamma(z_bz_c,z_bz_c^{-1},z_b^{-1}z_c,z_b^{-1}z_c^{-1})}\nonumber\\
\end{align}
Of course, writing down the index and evaluating it are two different things.  In what follows we will have more to say about both exact and approximate methods.

\subsection{Properties of Elliptic Gamma Functions}

The most obvious property of the elliptic gamma function is the exchange symmetry,
\be
\Gamma(u;p,q)=\Gamma(u;q,p).
\ee
This is manifest in the single-letter index \eqref{letter}.  This is just the statement that physics doesn't change if you stand on your head; the eigenvalues of $J_3$ and $-J_3$ are the same.  Another symmetry that can be seen from the single-letter index is 
\be\label{inversegamma}
\Gamma(u;p,q)=\Gamma(pqu^{-1};p,q)^{-1},
\ee
which follows simply from pulling a minus sign to the front in \eqref{letter}.  This overall minus gives the inverse of the gamma, and the $u\to pqu^{-1}$ is just from exchanging the roles (signs) of the two terms in \eqref{letter}.  This can be interpreted as the obvious physical statement that massive fields don't contribute to the index.  If a mass term can be written
\be
W\supset m\Phi\wt\Phi
\ee
then two fields transform as conjugates, and have $R$-charges that sum to two, so
\be
i_\Phi+i_{\wt\Phi}=\frac{(pq)^{r/2}\chi-(pq)^{(2-r)/2}\bar\chi}{(1-p)(1-q)}+\frac{(pq)^{(2-r)/2}\bar\chi-(pq)^{(2-[2-r])/2}\chi}{(1-p)(1-q)}=0.
\ee
Or equivalently, $\Gamma\left((pq)^{r/2}\eta\right)\Gamma\left((pq)^{(2-r)/2}\eta^{-1}\right)=1$ as in \eqref{inversegamma}.
Obviously a single field with $R$-charge 1 that transforms in a real representation ($\chi(u)=\bar\chi(u)=\chi(u^{-1})$) doesn't contribute either.  

Most of the other properties have less obvious physical significance\footnote{For a readable discussion of many properties, including a derivation of some intriguing modular properties see \cite{1999math7061F}}.  For example, consider the elliptic gamma function's defining equation,
\be
\Gamma(pu;p,q)=\t(u;q)\Gamma(u;p,q),\qquad\Gamma(qu;p,q)=\t(u;p)\Gamma(u;p,q).
\ee
This property is the analog of the famous equation $\Gamma(1+x)=x\Gamma(x)$ for the ordinary gamma function.  The physical content of the relation is not clear to me, but it can be used to extract physically meaningful data.  In fact, it is useful to define the following generalized theta functions.
\be\label{gentheta}
\t(u;p;q)_n=\frac{\Gamma(uq^n;p,q)}{\Gamma(u;p,q)}=\left\{\begin{matrix}\prod_{a=0}^{n-1}\t(uq^a;p)&\mbox{for $n>0$}\\1&\mbox{for $n=0$}\\\prod_{a=1}^{-n}\t(uq^{-a};p)^{-j}&\mbox{for $n<0$}\end{matrix}\right.
\ee
Thinking of $p$ as a unit-magnitude complex number, we can understand the nomenclature of the {\it quasiperiodicity relations},
\begin{align}
\t(p^Au;p;q)_C&=(\mbox{--}\,u)^{-AC}q^{-AC(C-1)/2}p^{-C A(A-1)/2}\t(u;p;q)_C,\label{quasi1}\\
\t(u;p;p^Bq)_C&=(\mbox{--}\,u)^{-BC(C-1)/2}q^{-BC(C-1)(2C-1)/6}p^{-BC(C-1)[B(2C-1)-3]/12}\t(u;p;q)_C.\label{quasi2}
\end{align}
These relations are used to simplify the total ellipticity conditions as defined in \cite{Spiridonov:2009za}.  Before giving the formal definition, however, let's consider a simple motivating example.  Consider a chiral superfield with zero $R$-charge but transforming under some global symmetry, so that it has an index of the form
\be
\prod_n\Gamma(y^{A_n}),\qquad y^{A_n}\equiv\prod_ay_a^{A_n^{(a)}}
\ee
Now consider choosing two of these fugacities and rescaling one by $p$ and the other by $q$.  Then using \eqref{gentheta} to relate this to the original index, one finds
\begin{align}\label{gammashift}
\prod_n\frac{\Gamma(p^{A_n^{(a)}}q^{A_n^{(b)}}y^{A_n})}{\Gamma(y^{A_n})}&=\prod_n\t(p^{A_n^{(a)}}y^{A_n};p;q)_{A_n^{(b)}}\t(y^{A_n};q,p)_{A_n^{(a)}}.\nonumber\\
&=(-1)^{f_-(A)}y^{f_y(A)}p^{f_p(A)}q^{f_q(A)}\prod_n\t(y^{A_n};q;p)_{A_n^{(a)}}\t(y^{A_n};p;q)_{A_n^{(b)}}
\end{align}
The second line follows from using \eqref{quasi1}.  The explicit forms of the exponents are the interesting part of this story.  For example, the coefficient of a fugacity $y_c$ is
\be
f_{y_c}(A)=-\sum_nA_n^{(a)}A_n^{(b)}A_n^{(c)}.
\ee
Though perhaps in an unfamiliar form, we have found the cubic invariant familiar from anomaly computations, $\Tr(T_\l^a\{T_\l^b,T_\l^c\})=T_3(\l)d^{abc}$.  For example, consider the two-index symmetric representation of $SU(3)$.
\be
\chi_{SU(3),\,\twosym{.35}}(y)=y_1^{-1}+y_2^{-1}+y_1y_2+y_1^2+y_2^2+y_1^{-2}y_2^{-2},\quad A_{\,\twosym{.35}}^{}=
\left(
\begin{matrix}
-1&0&1&2&0&-2\\
\,0&\!\!-1&1&0&2&-2
\end{matrix}
\right)
\ee
Choosing a non-zero combination, we find $\sum_{n=1}^6A_{\,\twosym{.3},n}^{(1)}A_{\,\twosym{.3},n}^{(1)}A_{\,\twosym{.3},n}^{(2)}=-7$.  To normalize to the fundamental, we also need 
\be
\chi_{SU(3),\Box}=y_1+y_2+y_1^{-1}y_2^{-1},\quad A_\Box=
\left(
\begin{matrix}
1&0&-1\\
0&1&-1
\end{matrix}
\right),
\ee
which gives $\sum_{i=1}^3A_{\Box,i}^{(1)}A_{\Box,i}^{(1)}A_{\Box,i}^{(2)}=-1$.  And so we find $T_3(\twosym{.4})/T_3(\fund{.5})=7$ in accord with $N+4$, the familiar result for the two-index symmetric tensor of $SU(N)$.

In more general theories one more ingredient is needed before defining total ellipticity.  For a field with a generic $R$-charge, there is an additional factor of the form $(pq)^{r/2}$.  We are told \cite{Spiridonov:2009za} to account for this by first defining
\be
u_0=(pq)^{1/B}.
\ee
We will then consider a rescaling of the form $q\to p^Bq$ to effect the desired change, $u_0\to pu_0$.  

Let's now start again from the beginning by defining the object of interest.  Given an equality of indices, $\I_E=\I_M$, we take the ratio of the integrands,
\be\label{delta1}
\Delta(u;p,q)=\frac{\Delta(u;p,q)_E}{\Delta(u;p,q)_M},\qquad\I_E=\int[dz]\Delta_E,\qquad\I_M=\int[d\tilde z]\Delta_M,
\ee
where $u$ stands in for all fugacities: global $(y)$, gauge $(z,\tilde z)$, and our special $R$-symmetry fugacity, $u_0$.  This ratio of integrands can be written in general as 
\be\label{delta2}
\Delta(u;p,q)=\prod_n\Gamma(u^{A_n})^{\e_n},\qquad u^{A_n}=\prod_au_a^{A_n^{(a)}}
\ee
The ranges of these products can be specified precisely in general, but it obvious enough in a given example how many gammas and how many parameters there are.  

Now we rescale some fugacity $u_a\ne u_0$ by $q$ and divide by the original function to eliminate the gammas, as in \eqref{gammashift}.  This new object is known as a {\it q-certificate}.
\be\label{qcert}
h_a(u;p;q)=\frac{\Delta(u_0,\dots,qu_a,\dots;p,q)}{\Delta(u;p,q)}=\prod_n\t(u^{A_n};p;q)_{A_n^{(a)}}^{\e_n},\qquad a\ne0
\ee
Total ellipticity is then defined as the invariance\footnote{In fact, it appears as though we should only demand invariance up to an overall sign.  See appendix \ref{ellapp} for further discussion of this point.} of the $q$-certificates under rescaling the $u$ by $p$, including a coupled $q$-$u_0$ rescaling.
\be\label{qcertinv}
h_a(u_0,\dots,pu_b,\dots;p;q)=h_a(pu_0,u_1,\dots;p;p^Bq)=h_a(u;p;q),\qquad a\ne0
\ee
This calculation is somewhat involved and some aspects of the prescription and its interpretation remain unclear to me, so I've included more details in appendix \ref{ellapp}.  Using \eqref{quasi1} and \eqref{quasi2} I find that \eqref{qcertinv} implies
\begin{align}
\sum_n\e_nA_n^{(a)}A_n^{(b)}A_n^{(c)}&=0,\label{ggg}\\
\sum_n\e_nA_n^{(a)}A_n^{(b)}\left(2\frac{A_n^{(0)}}B-1\right)&=0,\label{ggr}\\
\sum_n\e_nA_n^{(a)}\left(2\frac{A_n^{(0)}}B-1\right)^2&=0,\label{grr}\\
\sum_n\e_nA_n^{(a)}&=0.\label{g}
\end{align}
Recalling that $2A_n^{(0)}/B$ is the $R$-charge of the $n$-th chiral field (and $R-1$ gives the fermion's charge) we immediately recognize anomaly equations.  Since these equations must hold for all non-zero $a$, $b$, and $c$, we have pure gauge, pure global, and mixed anomaly constraints all compactly represented.  For example, in \eqref{ggg}, since the electric and magnetic gauge fugacities are independent---indeed, they are nothing but dummy variables---these anomalies must independently vanish, $\Tr G_EG_EG_E=\Tr G_MG_MG_M=0$.  The same applies to the linear term \eqref{g}.    For the pure global counterparts, we have 't Hooft anomaly matching constraints of the form $\Tr_E FFF=\Tr_M FFF$ and $\Tr_EF=\Tr_MF$.  These follow from the fact that an electric chiral superfield has $\e=+1$ and a magnetic chiral superfield has $\e=-1$ (see \eqref{delta1} and \eqref{delta2}).  From \eqref{ggg} we also have the constraints of the form $\Tr FGG=0$ that ensure that our flavor symmetries are non-anomalous.  And we have $\Tr FFG$=0, which is not independent.  

The equations involving the $R$-charges, \eqref{ggr} and \eqref{grr}, include the constraint that the $R$-symmetry be non-anomalous along with certain 't Hooft anomaly matching conditions.  We do not, however, have the full compliment.  The pure-$R$ 't Hooft anomalies, $\Tr U(1)_R$ and $\Tr U(1)_R^3$, are absent.  Given the lack of a zeroeth $q$-certificate \eqref{qcert} it's not surprising that they don't appear, but it's also not obvious to me how to extend this definition so that these constraints are included.  The naive guesses seem to fail.

\section{The Deconfined Two-Index Anti-symmetric Tensor}\label{antisec}

We will now consider the integral evaluation implied by the two-index anti-symmetric tensor deconfinement module as defined in \cite{Luty:1996cg}.  This provides a nice testing ground for some of the ideas and techniques discussed earlier.  We will also see in section \ref{ceht} that it proves to be a valuable key that can be used to unlock more intricate results.

The two-index anti-symmetric tensor of $SU(N)$ was the original subject of the Berkooz trick \cite{Berkooz:1995km}.  He ``deconfined'' the two-index tensor by taking it to be a bound state of fundamentals under a symplectic gauge group.  Identification of Seiberg dual theories then followed from the known dualities \cite{Seiberg:1994pq,Intriligator:1995ne}. In \cite{Berkooz:1995km} there was a single $Sp$-charged field, implying a non-trivial moduli space defined by the $D$-term constraint. This was accounted for with an IR superpotential proportional to Pf$A$.  Thanks to additional fields this is no longer the case for the deconfinement module of \cite{Luty:1996cg}, which is defined below.
\be\label{twoantiw}
W_{UV}=aX_1X_2+X_1X_1X_3,
\ee
\be\label{twoantitable}
\begin{tabular}{c|c||c|c|c|c}
&$Sp(N+K-4)$&$G_N$&$SU(K)$&$U(1)$&$U(1)_R$\\\hline
$a$&$\fund{.5}$&$\fund{.5}$&1&$e$&$r$\\\hline
$X_1$&$\fund{.5}$&1&$\fund{.5}$&$-Ne/K$&$(2-Nr)/K$\\\hline
$X_2$&1&$\overline{\fund{.5}}$&$\overline{\fund{.5}}$&$(N-K)e/K$&$[2(K-1)+(N-K)r]/K$\\\hline
$X_3$&1&1&$\twoantic{.5}$&$2Ne/K$&$2(K-2+Nr)/K$\\\hline\hline
$A$&$\cdot$&$\twoanti{.5}$&1&$2e$&$2r$\\
\end{tabular}
\ee
where $G_N$ is any of the classical Lie groups, $SU(N)$, $Sp(N)$ or $SO(N)$. It is left ungauged for now. Of course, $N+K$ must be even ($Sp(2)\cong SU(2)$ in this paper) and greater than 4, and when $G_N=Sp(N)$, $N$ and $K$ must be even. The superpotential is constructed to give mass to the singlets, $X_2$ and $X_3$, and the unwanted mesons, $aX_1$ and $X_1X_1$, upon confinement. There is a unique way to contract the indices to get singlets, so they are suppressed. One might notice similarities in the abelian charges in the table.  This is because we have taken the most general $R$-symmetry. It can be simplified considerably by taking the linear combination $U(1)_R^\prime=U(1)_R-\frac reU(1)$ (or equivalently, by choosing $r=0$).  $R$-symmetries will be kept general throughout for two reasons.  First, it is often useful to use the ambiguity to choose convenient\footnote{There is a particular form of convenience that can be given an explicit definition.  As can be seen from \eqref{letter}, when all $R$-charges are between 0 and 2, there is a well-defined expansion about $t=(pq)^{1/2}=0$.} charges.  Second, it is necessary to identify the particular linear combination of the symmetries that gives the superconformal $R$-symmetry if we want to interpret the index as the IR superconformal index.  

Without further ado we present the identity.  Defining $2M\equiv N+K-4$, we have
\be\label{ie=im1}
\I_{X_2}\I_{X_3}\int[dz]_{Sp(2M)}\I_a\I_{X_1}=\I_A,
\ee
where, assigning fugacities,
\be
U(1):y_1,\qquad G_N:y_2,\qquad SU(K):y_3
\ee
we have the following single-field contributions.
\begin{enumerate}
\item$G_N=SU(N)$
\begin{flalign}
{\cal I}_a&=\prod_{a=1}^N\prod_{b=1}^M\Gamma\big((pq)^{r/2}y_1^ey_{2,a}z_b^{\pm1}\big),&\\
{\cal I}_{X_1}&=\prod_{a=1}^K\prod_{b=1}^M\Gamma\big((pq)^{(2-Nr)/(2K)}y_1^{-Ne/K}y_{3,a}z_b^{\pm1}\big),&\\
{\cal I}_{X_2}&=\prod_{a=1}^{N}\prod_{b=1}^{K}\Gamma\big((pq)^{[K-1+(N-K)r/2]/K}y_1^{(N-K)e/K}y_{2,a}^{-1}y_{3,b}^{-1}\big),&\\
{\cal I}_{X_3}&=\prod_{1\le a<b\le K}\Gamma\big((pq)^{(K-2+Nr)/K}y_1^{2Ne/K}y_{3,a}^{-1}y_{3,b}^{-1}\big),&\\
{\cal I}_{A}&=\prod_{1\le a<b\le N}\Gamma\big((pq)^{r}y_1^{2e}y_{2,a}y_{2,b}\big),&
\end{flalign}
where $\prod_{a=1}^N y_{2,a}=\prod_{a=1}^Ky_{3,a}=1$ and $N+K$ is even.
\item$G_N=Sp(N=2n)$
\begin{flalign}
{\cal I}_a&=\prod_{a=1}^n\prod_{b=1}^M\Gamma\big((pq)^{r/2}y_1^ey_{2,a}^\pm z_b^\pm\big),&\\
{\cal I}_{X_1}&=\prod_{a=1}^K\prod_{b=1}^M\Gamma\big((pq)^{(2-Nr)/(2K)}y_1^{-Ne/K}y_{3,a}z_b^\pm\big),&\\
{\cal I}_{X_2}&=\prod_{a=1}^{n}\prod_{b=1}^{K}\Gamma\big((pq)^{[K-1+(N-K)r/2]/K}y_1^{(N-K)e/K}y_{2,a}^\pm y_{3,b}^{-1}\big),&\\
{\cal I}_{X_3}&=\prod_{1\le a<b\le K}\Gamma\big((pq)^{(K-2+Nr)/K}y_1^{2Ne/K}y_{3,a}^{-1}y_{3,b}^{-1}\big),&\\
{\cal I}_{A}&=\Gamma\big((pq)^{r}y_1^{2e}]^n\left(\prod_{1\le a<b\le n}\Gamma\big((pq)^{r}y_1^{2e}y_{2,a}^{\pm1}y_{2,b}^{\pm1}]\right)\prod_{a=1}^n\Gamma\big((pq)^{r}y_1^{2e}y_{2,a}^{\pm2}\big),&
\end{flalign}
where $\prod_{a=1}^Ky_{3,a}=1$ and $K$ is even.
\item$G_N=SO(N=2n)$
\begin{flalign}
{\cal I}_a&=\prod_{a=1}^n\prod_{b=1}^M\Gamma\big((pq)^{r/2}y_1^ey_{2,a}^\pm z_b^\pm\big),&\\
{\cal I}_{X_1}&=\prod_{a=1}^K\prod_{b=1}^M\Gamma\big((pq)^{(2-Nr)/(2K)}y_1^{-Ne/K}y_{3,a}z_b^\pm\big),&\\
{\cal I}_{X_2}&=\prod_{a=1}^{n}\prod_{b=1}^{K}\Gamma\big((pq)^{[K-1+(N-K)r/2]/K}y_1^{(N-K)e/K}y_{2,a}^\pm y_{3,b}^{-1}\big),&\\
{\cal I}_{X_3}&=\prod_{1\le a<b\le K}\Gamma\big((pq)^{(K-2+Nr)/K}y_1^{2Ne/K}y_{3,a}^{-1}y_{3,b}^{-1}\big),&\\
{\cal I}_{A}&=\Gamma\big((pq)^{r}y_1^{2e}]^n\prod_{1\le a<b\le n}\Gamma\big((pq)^{r}y_1^{2e}y_{2,a}^\pm y_{2,b}^\pm\big),&
\end{flalign}
where $\prod_{a=1}^Ky_{3,a}=1$ and $K$ is even.
\item$G_N=SO(N=2n+1)$
\begin{flalign}
{\cal I}_a&=\prod_{b=1}^M\Gamma\big((pq)^{r/2}y_1^ez_b^{\pm1}]\prod_{a=1}^N\Gamma\big((pq)^{r/2}y_1^ey_{2,a}^{\pm1}z_b^{\pm1}\big),&\\
{\cal I}_{X_1}&=\prod_{a=1}^K\prod_{b=1}^M\Gamma\big((pq)^{(2-Nr)/(2K)}y_1^{-Ne/K}y_{3,a}z_b^{\pm1}\big),&\\
{\cal I}_{X_2}&=\prod_{b=1}^{K}\Gamma\big((pq)^{[K-1+(N-K)r/2]/K}y_1^{(N-K)e/K} y_{3,b}^{-1}\big)&\\
&\hspace{1in}\times\prod_{a=1}^n\Gamma\big((pq)^{[K-1+(N-K)r/2]/K}y_1^{(N-K)e/K}y_{2,a}^{\pm1} y_{3,b}^{-1}\big),&\\
{\cal I}_{X_3}&=\prod_{1\le a<b\le K}\Gamma\big((pq)^{(K-2+Nr)/K}y_1^{2Ne/K}y_{3,a}^{-1}y_{3,b}^{-1}\big),&\\
{\cal I}_{A}&=\Gamma\big((pq)^{r}y_1^{2e}]^n\left(\prod_{1\le a<b\le n}\Gamma\big((pq)^{r}y_1^{2e}y_{2,a}^{\pm1} y_{2,b}^{\pm1}]\right)\prod_{a=1}^n\Gamma\big((pq)^{r}y_1^{2e}y_{2,a}^{\pm1}\big),&
\end{flalign}
where $\prod_{a=1}^Ky_{3,a}=1$ and $K$ is odd.
\end{enumerate}

I'll only prove the first of these identities.  In fact, there isn't any heavy lifting to do\footnote{I am grateful to Eric Rains for pointing me to his result.}.  The result follows from the integral identity \cite{2003math9252R} expressing the equality of the electric and magnetic indices for $Sp(N_c)$ SQCD \cite{Intriligator:1995ne}.  The relevant version of this identity (Corollary 3.2 of \cite{2003math9252R}) is the s-confining case, which can be summarized succinctly as follows.
\be\label{scontable}
\begin{tabular}{c|c||c|c}
&$Sp(2M)$&$SU(2M+4)$&$U(1)_R$\\\hline
$\Phi$&$\fund{.5}$&$\fund{.5}$&$\frac1{M+2}$\\\hline\hline
$M_{\Phi\Phi}$&$\cdot$&$\twoanti{.5}$&$\frac2{M+2}$
\end{tabular}
\ee
The confined theory also has the superpotential $W=\mbox{Pf}M_{\Phi\Phi}$.  Note that the $R$-charge is fixed in the gauged theory by the condition that it be non-anomalous, it maps to the IR through the interpretation of $M_{\Phi\Phi}$ as a bound state, and it is consistent with the superpotential.  
The integral identity can be written as
\be\label{spscon}
\int[dz]_{Sp(2M)}\prod_{a=1}^{2M+4}\prod_{b=1}^M\Gamma(v_az_b^{\pm1})=\prod_{1\le a<b\le 2M+4}\Gamma(v_av_b),\qquad \qquad\prod_{a=1}^{2M+4}v_a=pq
\ee
where $v_a=(pq)^{1/(2M+4)}y_a$ and $\prod_{a=1}^{2M+4}y_a=1$.  It is important to note that the integral is insensitive to particular choices of individual fugacities.  It is only the product that matters.  

To take advantage of \eqref{spscon}, we start by combining the integrand,
\be
\I_a\I_{X_1}=\prod_{a=1}^{N+K}\prod_{b=1}^M\Gamma(u_az_b^{\pm1})\qquad {\bf u}=\big((pq)^{r/2}y_1^e{\bf y_2},(pq)^{(2-Nr)/(2K)}y_1^{-Ne/K}\boldsymbol{y_3}\big)
\ee
Recall that $2M+4=N+K$, so we have reproduced the left side of \eqref{spscon}.  One can also easily verify that $\prod_{a=1}^{2M+4}u_a=pq$ as required.  Using \eqref{inversegamma} the remaining contributions are moved to the right side,
\begin{align}\label{ix2}
\I_{X_2}^{-1}&=\prod_{a=1}^N\prod_{b=1}^K\Gamma\big([(pq)^{[1-(N-K)r/2]/K)}y_1^{(K-N)e/K}y_{2,a}y_{3,b}\big)\nonumber\\
&=\prod_{a=1}^N\prod_{b=N+1}^{N+K}\Gamma(u_au_b)\\
\I_{X_3}^{-1}&=\prod_{1\le a<b\le K}\Gamma\big((pq)^{(2-Nr)/K}y_1^{-2Ne/K}y_{3,a}y_{3,a}\big)\nonumber\\
&=\prod_{N+1\le a<b\le N+K}\Gamma(u_au_b)
\end{align}
The last contribution also takes a simple form,
\be\label{iA}
\I_{A}=\prod_{1\le a<b\le N}\Gamma(u_au_b).
\ee
These three products combine to a single one, yielding the desired result,
\be\label{twoantiidentity}
\int[dz]_{Sp(2M)}\I_a\I_{X_1}=\int[dz]_{Sp(2M)}\prod_{a=1}^{2M+4}\prod_{b=1}^M\Gamma(u_az_b^{\pm1})=\prod_{1\le a<b\le 2M+4}\Gamma(u_au_b)=\I_{X_2}^{-1}\I_{X_3}^{-1}\I_A.
\ee

One of the most interesting aspects of the confining theory \eqref{twoantitable} and the associated integral identities is the free discrete parameter $K$, which allows for arbitrarily large gauge group rank in the UV theory and vanishes completely in the IR. The non-perturbative truncation of the chiral ring is a standard phenomenon \cite{Intriligator:1995au}.  What is interesting is not that the $SU(K)$-charged operators are eliminated from the chiral ring, but that the index doesn't have to be told that this symmetry is essentially fake; the fugacities associated with it simply cancel out.  This follows from the general result \eqref{twoantiidentity}, but it is amusing to observe it order by order in $t$.  

Below we present the result of a particular expansion with $N=5$, $e=1/2$, and $r=1/3$.  The $SU(5)$ characters\footnote{Of course, the description in terms of the anti-symmetric tensor is artificial here.  The symmetry is enhanced to $SU(N(N-1)/2)\times U(1)$ with our free fields transforming as a fundamental.  We have in mind exploiting the identity \eqref{ie=im1} in interacting cases where the identity of a two-index anti-symmetric tensor is unambiguous.} are denoted by $\chi_{(r_1,r_2,\dots)}$ where $r_n$ is the number of boxes in the $n$th row of the corresponding Young tableaux.
\begin{align}\label{scexpand}
\I_{UV}=\I_{IR}=&\,1\nonumber\\
&+y_1\chi_{(1,1,0,0,0)}t^{2/3}\nonumber\\
&+\left[-y_1^{-1}\chi_{(1,1,1,0,0)}+y_1^2(\chi_{(1,1,1,1,0)}+\chi_{(2,2,0,0,0)})\right]t^{4/3}\nonumber\\
&+(x+x^{-1})y_1\chi_{(1,1,0,0,0)}t^{5/3}\nonumber\\
&+\left[-1-\chi_{(2,1,1,1,0)}-\chi_{(2,2,1,0,0)}+y_1^3(\chi_{(2,2,1,1,0)}+\chi_{(3,3,0,0,0)})\right]t^2\nonumber\\
&+(x+x^{-1})(-y1^{-1}\chi_{(1,1,1,0,0)}+y1^2(\chi_{(1,1,1,1,0)}+\chi_{(2,1,1,0,0)}+\chi_{(2,2,0,0,0)})t^{7/3}\nonumber\\
&+\O(t^{8/3}).
\end{align}
The expansion is easily extended to much higher orders, but this will suffice.  Perhaps it is first worth saying something about how this result emerges from the UV index, since the footprints leading us here are rather well obscured.  For example, for the above case and $K=3$ the vanishing of the $R=1/9$ term is non-trivial:
\be
\I_{UV}\supset\frac{y_1^{-4/3}}2(\chi_{SU(3),\Box}(y_3)^2-\chi_{SU(3),\Box}(y_3^2)-2\chi_{SU(3),\twoanti{.35}}(y_3))t^{1/9}=0.
\ee
Similar but much more complicated identities come in at higher orders to yield the simple expansion \eqref{scexpand}.

Another interesting aspect of the expansion is that it is consistent with \eqref{scinterpret} and thus the interpretation of this index as a superconformal index (a theory of free fields is superconformal and the fields' superconformal $R$-charge is 2/3).  A prediction that follows from \eqref{scinterpret} that can be checked by eye is that the representations appearing at order $t^n$ must also appear at orders $t^{n+m}$ for $m=1,2,3\dots$  Of course, from the onset the free theory was really the only case in which we were confident that this would work, so given the integral identity \eqref{twoantiidentity}, this stops being interesting.

\section{CEHT Duality}\label{ceht}

I will now demonstrate that the index remains fixed throughout the chain of IR equivalent theories constructed in \cite{Craig:2011tx}.  The steps involved are deconfinement as in \eqref{twoantiidentity}, classic Seiberg duality, and $Sp$ s-confinement as in \eqref{spscon}.  As I mentioned earlier, all of these dualities are manifestations of a single type of transformation from the point of view of elliptic hypergeometric integrals.

\subsection{Electric Theory}

Our starting point is a four dimensional, $\N=1$, $SU(N)$ gauge theory with a two-index anti-symmetric tensor and fundamentals and anti-fundamentals as anomaly cancelation permits.  The theory is taken to have the superpotential,
\be
W_E=\wt QA\wt Q.
\ee
$2N-2$ anti-fundamentals, $Q$, couple the anti-symmetric tensor, $A$, and the number of fundamentals is taken to be $N_\Box=N+3$ \cite{Craig:2011tx}, leaving one additional anti-fundamental, $\wt P$.  These are simplifying choices.  See \cite{Craig:2011wj} for a discussion of the theory for all partitions.  The table:
\be\label{etable}
\begin{tabular}{c|c||c|c|c|c}
&$SU(N)$&$Sp(2N-2)$&$SU(N+3)$&$U(1)$&$U(1)_R$\cr\hline
$Q$&$\fund{.5}$&1&$\overline{\fund{.5}}$&$e$&$r$\cr\hline
$\wt Q$& $\overline{\fund{.5}}$& $\fund{.5}$ & 1 &$\tilde e$&$\tilde r$\cr\hline
$\wt P$&$\overline{\fund{.5}}$&1&1&$-(N+3)e-2\tilde e$&$4-(N+3)r-2\tilde r$\cr\hline
$A$&$\twoanti{.5}$&1&1&$-2\tilde e$&$2-2\tilde r$\cr
\end{tabular}
\ee
As before the symmetries to the right of the double line are global, and the $U(1)$'s shown are non-anomalous.  From \eqref{word} we can quickly right down the index for this theory.
\be
\I_E=\int[dz]_{SU(N)}\I_Q\I_{\wt Q}\I_{\wt P}\I_A
\ee
Making the fugacity assignments,
\be
U(1):y_1,\qquad Sp(2N-2):y_2,\qquad SU(N+3):y_3,
\ee
the single field contributions to the integrand are
\begin{flalign}
\qquad\I_Q&=\prod_{a=1}^{N+3}\prod_{b=1}^N\Gamma\big((pq)^{r/2}y_1^ey_{3,a}^{-1}z_b\big),&\\
\I_{\wt Q}&=\prod_{a=1}^{N-1}\prod_{b=1}^N\Gamma\big((pq)^{\tilde r/2}y_1^{\tilde e}y_{2,a}^{\pm1}z_b^{-1}\big),&
\end{flalign}
\begin{flalign}
\qquad\I_{\wt P}&=\prod_{a=1}^N\Gamma\big((pq)^{[4-(N+3)r-2\tilde r]/2}y_1^{-(N+3)e-2\tilde e}z_a^{-1}\big),&\\
\I_A&=\prod_{1\le a<b\le N}\Gamma\big((pq)^{1-\tilde r}y_1^{-2\tilde e}z_az_b\big).&
\end{flalign}

\subsection{Deconfined Theory}

The elimination of the field $A$ follows immediately from the module constructed in \cite{Luty:1996cg} and discussed in section \ref{antisec}.  Taking $2M=N+K-4$ as before,
we have\footnote{I apologize for the notational failures in this paper, including the recycling of certain parameters like $e$ and $r$.} \cite{Craig:2011tx}
\be\label{dtable}
\begin{tabular}{c|c|c||c|c|c|c|c}
&\!\!$Sp(2M)$\!\!&\!\!$SU\hspace{-.07em}(\hspace{-.07em}N)$\!\!&\!\!$SU\hspace{-.07em}(\hspace{-.07em}K)$\!\!&$\!\!Sp(2N\!\!-\!2)$\!\!&\!\!$SU\hspace{-.07em}(\hspace{-.07em}N\!\!+\!3)$\!\!&$U(1)$&$U(1)_R$\cr\hline
\!\!\!$Q$\!\!&1&$\fund{.5}$&1&1&$\overline{\fund{.5}}$&$e$&$r$\cr\hline
\!\!\!$\wt Q$\!\!&1&$\overline{\fund{.5}}$&1&$\fund{.5}$&1&$\tilde e$&$\tilde r$ \cr\hline
\!\!\!$\wt P$\!\!&1&$\overline{\fund{.5}}$&1&1&1&$\!\hspace{-.07em}-(N\!\hspace{-.07em}+\!3)e\!-\!2\tilde e$\hspace{-.07em}\!&$4\!-\!(N\!+\!3)r\!-\!2\tilde r$\cr\hline
\!\!\!$a$\!\!&$\fund{.5}$&$\fund{.5}$&1&1&1&$-\tilde e$&$1-\wt r$\cr\hline
\!\!\!$X_1$\!\!&$\fund{.5}$&1&$\fund{.5}$&1&1&$N\tilde e/K$&$(2\!-\!N\!+\!N\tilde r)/K$\cr\hline
\!\!\!$X_2$\!\!&1&$\overline{\fund{.5}}$&$\overline{\fund{.5}}$&1&1&\!\!$(\hspace{-.07em}K\!\hspace{-.07em}-\!N)\tilde e/K$\!\!&$\!\!(N\!+\!K\!\hspace{-.07em}-\!2\!-\![N\!-\!K]\tilde r)/K$\!\!\cr\hline
\!\!\!$X_3$\!\!&1&1&$\overline{\twoanti{.5}}$&1&1&$-2N\tilde e/K$&$2(N\!+\!K\!-\!2\!-\!N\tilde r)/K$
\end{tabular}
\ee
and the superpotential is
\begin{align}
W_D&=\wt Qaa\wt Q+aX_1X_2+X_1X_1X_3
\end{align}
The index of this theory is identical to the electric index of the last subsection by \eqref{twoantiidentity}.  There is nothing further to prove.  For completeness it is written out below.  
\be\label{ie=id}
\I_E=\I_D=\I_{X_3}\int[dz_1]_{SU(N)}\I_{Q}\I_{\wt Q}\I_{\wt P}\I_{X_2}\int[dz_2]_{Sp(2M)}\I_a\I_{X_1}
\ee
Assigning fugacities as
\be
U(1):y_1,\quad Sp(2N-2):y_2,\quad SU(N+3):y_3,\quad SU(K):y_4,\quad SU(N):z_1,\quad Sp(2M):z_2,
\ee
the pieces of the index in \eqref{ie=id} are
\begin{flalign}
\qquad\I_Q&=\prod_{a=1}^N\prod_{b=1}^{N+3}\Gamma\big((pq)^{r/2}y_1^{e}z_{1,a}y_{3,b}^{-1}\big),&\label{IQ}\\
\I_{\wt Q}&=\prod_{a=1}^N\prod_{b=1}^{N-1}\Gamma\big((pq)^{\tilde r/2}y_1^{\tilde e}z_{1,a}^{-1}y_{2,b}^{\pm1}\big),&\\
\I_{\wt P}&=\prod_{a=1}^N\Gamma\big((pq)^{[4-(N+3)r-2\tilde r]/2}y_1^{-(N+3)e-2\tilde e}z_{1,a}^{-1}\big),&\\
\I_{a}&=\prod_{a=1}^N\prod_{b=1}^M\Gamma\big((pq)^{(1-\tilde r)/2}y_1^{-\tilde e}z_{1,a}z_{2,b}^{\pm1}\big),&\label{Ia}\\
\I_{X_1}&=\prod_{a=1}^K\prod_{b=1}^M\Gamma\big((pq)^{(2-N+N\tilde r)/(2K)}y_1^{N\tilde e/K}y_{4,a}z_{2,b}^{\pm1}\big),&\\
\I_{X_2}&=\prod_{a=1}^N\prod_{b=1}^K\Gamma\big((pq)^{[N+K-2-(N-K)\tilde r]/(2K)}y_1^{(K-N)\tilde e/K}z_{1,a}^{-1}y_{4,b}^{-1}\big),&\\
\I_{X_3}&=\prod_{1\le a<b\le K}\Gamma\big((pq)^{(N+K-2-N\tilde r)/K}y_1^{-2N\tilde e/K}y_{4,a}^{-1}y_{4,b}^{-1}
\big).&
\end{flalign}

\subsection{Deconfined Magnetic Theory}

Assuming now that the strong coupling scale of the $SU(N)$ gauge group is greater than that of the $Sp(2M)$ gauge group, we dualize according to the standard prescription.  It is essentially just $SU(N)$ SQCD with $N_f\equiv2N+K-1$ flavors.  This basic duality can be summarized as follows \cite{Seiberg:1994pq}.
\begin{flalign}\label{suseiberg}&\quad
\begin{tabular}{c|c||c|c|c|c}
&$SU(N)$&$SU(N_f)_1$&$SU(N_f)_2$&$U(1)_B$&$U(1)_R$\\\hline
$\Phi$&$\fund{.5}$&$\fund{.5}$&1&1&$1-\frac N{N_f}$\\\hline
$\wt \Phi$&$\overline{\fund{.5}}$&1&$\overline{\fund{.5}}$&$-1$&$1-\frac N{N_f}$
\end{tabular}\quad,\qquad W_E=0&
\end{flalign}
\begin{flalign}&\quad
\begin{tabular}{c|c||c|c|c|c}
&$SU(N_f-N)$&$SU(N_f)_1$&$SU(N_f)_2$&$U(1)_B$&$U(1)_R$\\\hline
$\phi$&$\fund{.5}$&$\overline{\fund{.5}}$&1&$\frac N{N_f-N}$&$\frac N{N_f}$\\\hline
$\tilde\phi$&$\overline{\fund{.5}}$&1&$\fund{.5}$&$-\frac N{N_f-N}$&$\frac N{N_f}$\\\hline
$M_{\Phi\wt\Phi}$&1&$\fund{.5}$&$\overline{\fund{.5}}$&0&$2(1-\frac N{N_f})$
\end{tabular}\quad,\qquad W_M=\phi M\tilde\phi&
\end{flalign}

To exploit this result we write the $SU(N)$ fundamental contributions \eqref{IQ} and \eqref{Ia} as if they comprised a single $SU(N_f)_1$ fundamental.
\be\label{sqcdindex}
\I_Q\I_a=\prod_{a=1}^{N_f}\prod_{b=1}^N\Gamma\big((pq)^{(1-N/N_f)/2}\tilde y_1\tilde y_{2,a}z_{1,b}\big)
\ee
The new parameters are related to the old by the definition,
\be\label{sqcdmatching1}
(pq)^{(1-N/N_f)/2}\tilde y_1 \boldsymbol{\tilde y_2}=\left((pq)^{r/2}y_1^e\boldsymbol{y_3^{-1}},(pq)^{(1-\tilde r)/2}y_1^{-\tilde e}\boldsymbol{z_2},(pq)^{(1-\tilde r)/2}y_1^{-\tilde e}\boldsymbol{z_2^{-1}}\right).
\ee
The form in \eqref{sqcdindex} is exactly that of SQCD with $\wt y_1$ playing the role of the baryon number fugacity and $\tilde y_2$ that associated with $SU(N_f)_1$. 
However, we must demand that $\prod_{a=1}^{N_f} \tilde y_{2,a}=1$ (the generators of $SU(N_f)_1$ are traceless) in \eqref{sqcdmatching1}.  This gives an equation for $\tilde y_1(y_1,pq)$.  Now we do the same thing for the anti-fundamentals.  We write
\be
\I_{\wt Q}\I_{\wt P}\I_{X_2}=\prod_{a=1}^{N_f}\prod_{b=1}^N\Gamma\big((pq)^{(1-N/N_f)/2}\tilde y_1^{-1}\tilde y_{3,a}^{-1}z_{1,b}^{-1}\big),
\ee
where $\tilde y_3$ would be the fugacity for the $SU(N_f)_2$ symmetry.  The old and new parameters are related as follows.
\begin{align}\label{sqcdmatching2}
(pq)^{(1-N/N_f)/2}\tilde y_1^{-1} \boldsymbol{\tilde y_3^{-1}}
=\big((pq)^{\tilde r/2}y_1^{\tilde e}\boldsymbol{y_2},(pq)^{\tilde r/2}y_1^{\tilde e}\boldsymbol{y_2^{-1}}&,(pq)^{[4-(N+3)r-2\tilde r]/2}y_1^{-(N+3)e-2\tilde e},\nonumber\\
&(pq)^{[N+K-2-(N-K)\tilde r]/(2K)}y_1^{(K-N)\tilde e/K}\boldsymbol{y_4^{-1}}\big).
\end{align}
While we have introduced new parameters, the condition $\prod_{a=1}^{N_f}\tilde y_{3,a}=1$ is not trivially satisfied.  This is again a function only of $y_1$, $\tilde y_1$, and $pq$.  One does indeed find that both \eqref{sqcdmatching1} and \eqref{sqcdmatching2} imply 
\be
\tilde y_1=(pq)^{[-3+(N+3)r-2M\tilde r]/(2N_f)}y_1^{[(N+3)e-2M\tilde e]/N_f}
\ee
So we see that we can directly apply the result of \cite{2003math9252R} and we have the explicit mapping of parameters.  By dualizing and then rewriting in terms of the original variables, we find the desired result,
\begin{align}\label{idm}
I_E\!=\!I_D\!&=\!I_{DM}\nonumber\\
&=\!\I_{X_3}\I_{M_{Q\wt Q}}\I_{M_{Q\wt P}}\I_{M_{Q\wt X_2}}\!\!\int\![dz_1]_{SU(N_f-N)}\I_q\I_{\tilde q}\I_{\tilde p}\I_\a\I_{x_2}\!\!\int\![dz_2]_{Sp(2M)}\I_\a\I_{X_1}\I_{M_{a\wt Q}}\I_{M_{a\wt P}}\I_{M_{aX_2}}.
\end{align}
The fields and the superpotential of this theory are shown below.  The charges have grown unwieldy, so they are mostly written in terms of the electric theory charges.  This description can be simplified by integrating out the massive fields.  This is automatic in the index; $\I_{M_{a\wt Q}}=\I_{M_{aX2}}\I_{X_1}=1$.  The other single-field contributions to the integrand are
\begin{flalign}
\qquad\I_q&=\prod_{a=1}^{N+3}\prod_{b=1}^{N_f-N}\Gamma\big((pq)^{r_q/2}y_1^{e_q}y_{3,a}z_{1,b}\big),&\\
\I_{\tilde q}&=\prod_{a=1}^{N-1}\prod_{b=1}^{N_f-N}\Gamma\big((pq)^{r_{\tilde q}/2}y_1^{e_{\tilde q}}y_{2,a}^{\pm1}z_{1,b}^{-1}\big),&\\
\I_{\tilde p}&=\prod_{a=1}^{N_f-N}\Gamma\big((pq)^{(2-r-r_q-r_{\wt P})/2}y_1^{-e-e_q-e_{\wt P}}z_{1,a}^{-1}\big),&\\
\I_{\a}&=\prod_{a=1}^{N_f-N}\prod_{b=1}^{M}\Gamma\big((pq)^{(1-r_{\tilde q})/2}y_1^{-e_{\tilde p}}z_{1,a}z_{2,a}^{\pm1}\big),&\\
\I_{x_2}&=\prod_{a=1}^{K}\prod_{b=1}^{N_f-N}\Gamma\big((pq)^{(2-r-r_q-r_{X_2})/2}y_1^{-e-e_q-e_{X_2}}y_{4,a}z_{1,b}^{-1}\big),&
\end{flalign}
\begin{flalign}
\qquad
\I_{X_3}&=\prod_{1\le a<b\le K}\Gamma\big((pq)^{r_{X_3}/2}y_1^{e_{X_3}}y_{4,a}^{-1}y_{4,b}^{-1}\big),&\\
\I_{M_{Q\wt Q}}&=\prod_{a=1}^{N-1}\prod_{b=1}^{N+3}\Gamma\big((pq)^{(r+\tilde r)/2}y_1^{e+\tilde e}y_{2,a}^{\pm1}y_{3,b}^{-1}\big),&\\
\I_{M_{Q\wt P}}&=\prod_{a=1}^{N+3}\Gamma\big((pq)^{(r+\wt P)/2}y_1^{e+e_{\wt P}}y_{3,a}^{-1}\big),&\\
\I_{M_{QX_2}}&=\prod_{a=1}^{N+3}\prod_{b=1}^{K}\Gamma\big((pq)^{(r+r_{X_2})/2}y_1^{e+e_{X_2}y_{3,a}^{-1}y_{4,b}^{-1}}\big),&\\
\I_{M_{a\wt P}}&=\prod_{a=1}^{M}\Gamma\big((pq)^{(r_a+r_{\wt P})/2}y_1^{e_a+e_{\wt P}}z_{2,a}^{\pm1}\big),&
\end{flalign}
where $\prod_{a=1}^{N+3}y_{3,a}=\prod_{a=1}^{K}y_{4,a}=\prod_{a=1}^{N_f-N}z_{1,b}=1$, and $N_f=2N+K-1$.
This is the index for the following theory \cite{Craig:2011tx}.
\be\label{dmtable}
\begin{tabular}{c|c|c||c|c|c|c|c}
&\!\!$Sp(2M)$\!\!&\!\!$SU(N_f\!-\!N)$\!\!&\!\!$SU(K)$\!\!&\!\!$Sp(2N\!-\!2)$\!\!&\!\!$SU(N\!+\!3)$\!\!&$U(1)$&$U(1)_R$\\\hline
\!\!\!$q$&1&$\fund{.5}$&1&1&${\fund{.5}}$&\!\!--\,$\frac{(K-4)e+2M\tilde e}{N+K-1}$\!\!&$\!\!\frac{-(K-4)r+2M(1-\tilde r)}{N+K-1}$\!\!\\\hline
\!\!\!$\tilde q$&1& $\overline{\fund{.5}}$&1&$\fund{.5}$&1&--\,$\frac{(N+3)e+3\tilde e}{N+K-1}$&\!\!$\frac{N+K+2-(N+3)r-3\tilde r}{N+K-1}$\!\!\\\hline
\!\!\!$\tilde p$&1&$\overline{\fund{.5}}$&1&1&1&\!\!--\,$e$\,--\,$e_q$\,--\,$e_{\wt P}$\!\!&\!\!\!2\,--\,$r$\,--\,$r_q$\,--\,$r_{\wt P}$\!\!\\\hline
\!\!\!$\a$&$\fund{.5}$&$\fund{.5}$&1&1&1&--\,$e_{\tilde q}$&1\,--\,$r_{\tilde q}$\\\hline
\!\!\!$X_1$&$\fund{.5}$&1&$\fund{.5}$&1&1&$e_{X_1}$&$r_{X_1}$\\\hline
\!\!\!$x_2$&1&$\overline{\fund{.5}}$&$\fund{.5}$&1&1&\!\!--\,$e$\,--\,$e_q$\,--\,$e_{X_2}$\!\!&\!\!\!2\,--\,$r$\,--\,$r_q$\,--\,$r_{X_2}$\!\!\\\hline
\!\!\!$X_3$&1&1&$\overline{\twoanti{.5}}$&1&1&$e_{X_3}$&$r_{X_3}$\\\hline
\!\!\!$M_{Q\wt Q}$\!\!&1&1&1&$\fund{.5}$&$\overline{\fund{.5}}$&$e+\tilde e$&$r+\tilde r$\\\hline
\!\!\!$M_{Q\wt P}$\!\!&1&1&1&1&$\overline{\fund{.5}}$&$e+e_{\wt P}$&$r+r_{\wt P}$\\\hline
\!\!\!$M_{QX_2}$\!\!&1&1&$\overline{\fund{.5}}$&1&$\overline{\fund{.5}}$&$e+e_{X_2}$&$r+r_{X_2}$\\\hline
\!\!\!$M_{a\wt Q}$\!\!&$\fund{.5}$&1&1&$\fund{.5}$&1&0&1\\\hline
\!\!\!$M_{a\wt P}$\!\!&$\fund{.5}$&1&1&1&1&$e_a+e_{\wt P}$&$r_a+r_{\wt P}$\\\hline
\!\!\!$M_{aX_2}$\!\!&$\fund{.5}$&1&$\overline{\fund{.5}}$&1&1&--\,$e_{X_1}$&2\,--\,$r_{X_1}$\\
\end{tabular}
\ee
\begin{align}
W_{DM}&=M_{a\wt Q}M_{a\wt Q}+M_{aX_2}X_1+X_1X_1X_3+\boldsymbol{q}\cdot\boldsymbol{M}\cdot\boldsymbol{\tilde q}\nonumber\\
\boldsymbol{M}&=\left(\begin{matrix}M_{Q\wt Q}&M_{Q\wt P}&M_{QX_2}\\M_{a\wt Q}&M_{a\wt P}&M_{aX_2}\end{matrix}\right),\qquad\boldsymbol{q}=(\,q\,\,\,\a),\qquad\boldsymbol{\tilde q}^T=(\,\tilde q\,\,\,\,\tilde p\,\,\,x_2)
\end{align}

Integrating out the heavy fields leaves
\be\label{wdm}
W_{DM}\to\tilde q\a\a\tilde q+x_2\a\a x_2X_3+qM_{Q\wt Q}\tilde q+qM_{Q\wt P}\tilde p+qM_{QX_2}x_2+\a M_{a\wt P}\tilde p.
\ee

\subsection{Magnetic Theory}\label{cehtmag}

After integrating out the massive fields, $X_1$, $M_{a\wt Q}$ and $M_{aX_2}$, the symplectic factor of the gauge group is left with just $N+K$ flavors, which is exactly the s-confining case \cite{Intriligator:1995ne}.  This was discussed in section \ref{antisec}; see \eqref{scontable}.  
 
Recall that in our case $2M=N+K-4$.  To interpret our fields $\a$ and $M_{a\wt P}$ as an $SU(2M+4)$ fundamental with the appropriate $R$-charge requires the identity
\be
\left((pq)^{(1-r_{\tilde q})/2}y_1^{-e_q}\right)^{N+K-1}(pq)^{(r_a+r_{\wt P})/2}y_1^{e_a+e_{\wt P}}=pq,
\ee
which can be verified from the tables \eqref{dtable} and \eqref{dmtable}.  There were no free parameters.  The exponents of $pq$ and $y_1$ both match automatically for all $r$, $\tilde r$, $e$, and $\tilde e$. 
This allows us to perform the integral in \eqref{idm} over the $Sp(2M)$, leaving
\be
\I_E=\I_D=\I_{DM}=\I_M=\I_{X_3}\I_{M_{Q\wt Q}}\I_{M_{Q\wt P}}\I_{M_{QX_2}}\int[dz]_{SU(N+K-1)}\I_q\I_{\tilde q}\I_{x_2}\I_{\cal A},
\ee
where ${\cal A}\sim\a\a$ represents the new anti-symmetric meson.  From \eqref{wdm} we see that the $\a$-$M_{a\wt P}$ bound state gets a mass along with field $\tilde p$.  As always, this is automatically accounted for in the index.  The only new non-trivial contribution to the index is
\be
\I_{\cal A}=\prod_{1\le a<b\le N+K-1}\Gamma\big((pq)^{1-r_{\tilde q}}y_1^{-2e_{\tilde q}}z_az_b\big)
\ee
This concludes the index version of the CEHT story \cite{Craig:2011tx}.  We have arrived at the rather economical theory,
\be\label{mtable}
\begin{tabular}{c|c||c|c|c|c|c}
&\!\!$SU(N\!+\!K$\,--\,$1)$\!\!&\!\!$SU(K)$\!\!&\!\!$Sp(2N\!-\!2)$\!\!&\!\!$SU(N\!+\!3)$\!\!&$U(1)$&$U(1)_R$\\\hline
\!\!\!$q$&$\fund{.5}$&1&1&${\fund{.5}}$&\!\!--\,$\frac{(K-4)e+2M\tilde e}{N+K-1}$\!\!&$\!\!\frac{-(K-4)r+2M(1-\tilde r)}{N+K-1}$\!\!\\\hline
\!\!\!$\tilde q$& $\overline{\fund{.5}}$&1&$\fund{.5}$&1&--\,$\frac{(N+3)e+3\tilde e}{N+K-1}$&\!\!$\frac{N+K+2-(N+3)r-3\tilde r}{N+K-1}$\!\!\\\hline
\!\!\!$x_2$&$\overline{\fund{.5}}$&$\fund{.5}$&1&1&\!\!--\,$e_q$\,--\,$e$\,--\,$e_{X_2}$\!\!&\!\!\!2\,--\,$r_q$\,--\,$r$\,--\,$r_{X_2}$\!\!\\\hline
\!\!\!${\cal A}$&$\twoanti{.5}$&1&1&1&--\,$2e_{\tilde q}$&2\,--\,$2r_{\tilde q}$\\\hline
\!\!\!$X_3$&1&$\overline{\twoanti{.5}}$&1&1&$e_{X_3}$&$r_{X_3}$\\\hline
\!\!\!$M_{Q\wt Q}$\!\!&1&1&$\fund{.5}$&$\overline{\fund{.5}}$&$e+\tilde e$&$r+\tilde r$\\\hline
\!\!\!$M_{Q\wt P}$\!\!&1&1&1&$\overline{\fund{.5}}$&$e+e_{\wt P}$&$r+r_{\wt P}$\\\hline
\!\!\!$M_{QX_2}$\!\!&1&$\overline{\fund{.5}}$&1&$\fund{.5}$&$e+e_{X_2}$&$r+r_{X_2}$\\
\end{tabular}.
\ee

\section{Conclusions}

There are two lessons from the models that we have studied that are worth reemphasizing.  First, different Seiberg duality systems need not derive from distinct integral transformations.  Even the highly exotic-looking duality between \eqref{etable} and \eqref{mtable} follows from the same integral transformations relating the electric and magnetic theories in vanilla SQCD.  This is encouraging.  If this were not true---if every dual system generated a new integral transformation---there would be little hope for interpreting the symmetries of the integrals as symmetries of Nature.  Second, we have seen that the index is not as stupid as one might have thought.  The built-in insensitivity to the non-perturbatively eliminated symmetry of \eqref{mtable} suggests that there is a sense in which the index ``knows'' about instantons and moduli spaces.

There are many interesting avenues for future work.  To name a few...
\begin{enumerate}
\item All of the proposed new dualities \cite{Spiridonov:2009za,Spiridonov:2011hf} should be examined.  
\item The conjectured integral identities following from known dualities \cite{Spiridonov:2009za,Spiridonov:2011hf} should be proven to see what, if any, fundamentally new integral transformations are at work.  
\item The conditions of total ellipticity should be reconsidered to see if a more general symmetry exists, as expected from the absence of certain 't Hooft anomaly matching conditions.  
\item The definition of the index and its interpretation should be placed on firmer theoretical foundations.  
\item The standing of index equality as a discriminant of duality should be rigorously evaluated.
\end{enumerate}

\section*{Acknowledgments}

I am grateful for the hospitality of the SLAC theory group and the Cornell University physics department, where this work was initiated.  I benefitted from many probing discussions with physicists and mathematicians at IAS, IPMU, and UCSD.  I should mention in particular Scott Carnahan, Thomas Dumitrescu, Alexander Getmanenko, Daniel Green, Jonathan Heckman, and Patipan Uttayarat.  I am especially indebted to Guido Festuccia, David Poland, Shlomo Razamat, and Ken Intriligator for extended and illuminating discussions, and to Eric Rains and Grisha Vartanov for authoring what are probably the two most time- and effort-saving emails I have ever received.  This research was supported by World Premier International Research Center Initiative (WPI Initiative), MEXT, Japan.

\appendix

\section{Group Theory Data}\label{grouptheoryapp}

There are three types of group theory data employed in this work: (1) eigenvalues of quadratic and cubic Casimir operators (the numbers that appear in anomalies), (2) characters of representations of Lie groups, and (3) measures on group space.  I'll discuss them in turn.

\subsection{Casimirs}

There are general formulas for eigenvalues of Casimir operators, but we only encountered four types of representation in this work\footnote{Other representations are rarely interesting in this context because gauge theories cease to be parametrically asymptotically free.}: fundamental, two-index anti-symmetric (adjoint of $SO$), two-index symmetric (adjoint of $Sp$), and the $SU$ adjoint and anti-fundamental ($SO$ and $Sp$ representations are real), so we will just list what we need.  The numbers of interest are defined as follows.
\be
\Tr( T_\l^aT_\l^b) =T_2(\l)\d^{ab},\qquad \Tr (T_\l^a\{T_\l^b,T_\l^c\})=T_3(\l)d^{abc}.
\ee
For $Sp(N)$ and $SO(N)$ we have 
\be
\frac{T_2(\twoanti{.4})}{T_2(\fund{.4})}=N-2,\qquad \frac{T_2(\twosym{.4})}{T_2(\fund{.4})}=N+2,\qquad T_3(\l)=0.
\ee
For $SU(N)$ we have
\be
\frac{T_2(\twoanti{.4})}{T_2(\fund{.4})}=N-2,\qquad \frac{T_2(\twosym{.4})}{T_2(\fund{.4})}=N+2,\qquad\frac{T_3(\twoanti{.4})}{T_3(\fund{.4})}=N-4,\qquad \frac{T_3(\twosym{.4})}{T_3(\fund{.4})}=N+4,
\ee
\be
\frac{T_2(\overline{\fund{.4}})}{T_2(\fund{.4})}=1,\qquad \frac{T_2(Adj)}{T_2(\fund{.4})}=2N,\qquad\frac{T_3(\overline{\fund{.4}})}{T_3(\fund{.4})}=-1,\qquad T_3(Adj)=0.
\ee
And, of course, the dimensions of the representations are dim$(\fund{.4})=$dim$(\overline{\fund{.4}})=N$, dim$(\twoanti{.4})=N(N-1)/2$, dim$(\twosym{.4})=N(N+1)/2$, and dim$(Adj)=N^2-1$.

\subsection{Characters}

Since characters of representations of Lie groups do not appear to be commonly used by particle physicists, I'll give a very brief introduction.  They are defined as a trace of a group element, $\Tr e^{i\a_a H_a}$, where $a$ runs over the rank of the group and the $H_a$ are Cartan generators in the representation of interest.  This is easier to digest through examples.  Let's consider $SU(2)$, where the Cartan generator for the spin-$j$ representation is given by,
\be
H_j=\mbox{diag}(j,j-1,\dots,-j+1,-j)
\ee
A precise formula for this example and our normalization is
\be
\chi_{SU(2),j}(u)=\Tr e^{2H_j\ln u}
\ee
For example, consider $j=3/2$.  In this case the character is found to be
\begin{align}
\chi_{SU(2),3/2}(u)
&=\Tr\exp\mbox{diag}(3\ln u,\ln u,-\ln u,-3\ln u)\nonumber\\
&=u^3+u+\frac1u+\frac1{u^3}
\end{align}
For $SU(3)$, Wikipedia's convention for the fundamental Cartan generators is
\be
H_{\Box,1}=\left(
\begin{matrix}
1&0&0\\
0&-1&0\\
0&0&0
\end{matrix}
\right)
\qquad
H_{\Box,2}=\frac1{\sqrt3}
\left(
\begin{matrix}
1&0&0\\
0&1&0\\
0&0&-2
\end{matrix}
\right)
\ee
Conventionally normalized characters are then given by choosing $\a$ in $e^{i\a H}$ to be 
\be
\a=-i\left(\frac12\ln(u_1/u_2),\frac{\sqrt3}2\ln(u_1u_2)\right)
\ee
In particular, the fundamental character becomes $\chi_{SU(3),\Box}=\Tr e^{i\a H_\Box}=u_1+u_2+u_1^{-1}u_2^{-1}$.  

Simple general formulas for the characters are known, but we will just list the ones that we use. 

\noindent For $SU(N)$ 
\begin{flalign}
\qquad\chi_{SU(N),\,\fund{.4}}(u)&=\sum_{a=1}^Nu_a,&\\
\chi_{SU(N),\,\overline{\fund{.4}}}(u)&=\sum_{a=1}^Nu_a^{-1},&\\
\chi_{SU(N),Adj}(u)&=-1+\!\!\sum_{1\le a,b\le N}\!u_au_b^{-1},&\\
\chi_{SU(N),\twoanti{.3}}(u)&=\sum_{1\le a<b\le N}\!u_au_b,&
\end{flalign}
where we must also impose $\prod_{a=1}^Nu_a=1$

\noindent For $Sp(2N)$
\begin{flalign}
\qquad\chi_{Sp(2N),\,\fund{.4}}(u)&=\sum_{a=1}^N(u_a+u_a^{-1}),&\\
\chi_{Sp(2N),\twoanti{.3}}(u)&=N-1+\!\!\sum_{1\le a<b\le N}(u_au_b+u_au_b^{-1}+u_a^{-1}u_b+u_a^{-1}u_b^{-1}),&\\
\chi_{Sp(2N),\,\twosym{.3}}(u)&=N+\sum_{a=1}^N(u_a^2+u_a^{-2})+\!\!\sum_{1\le a<b\le N}(u_au_b+u_au_b^{-1}+u_a^{-1}u_b+u_a^{-1}u_b^{-1}).&
\end{flalign}

\noindent For $SO(2N)$
\begin{flalign}
\qquad\chi_{SO(2N),\,\fund{.4}}(u)&=\sum_{a=1}^N(u_a+u_a^{-1}),&\\
\chi_{SO(2N),\twoanti{.3}}(u)&=N+\!\!\sum_{1\le a<b\le N}(u_au_b+u_au_b^{-1}+u_a^{-1}u_b+u_a^{-1}u_b^{-1}).&
\end{flalign}

\noindent For $SO(2N+1)$
\begin{flalign}
\qquad\chi_{SO(2N+1),\,\fund{.4}}(u)&=1+\sum_{a=1}^N(u_a+u_a^{-1}),&\\
\chi_{SO(2N+1), \twoanti{.3}}(u)&=N+\sum_{a=1}^N(u_a+u_b^{-1})+\!\!\sum_{1\le a<b\le N}(u_au_b+u_au_b^{-1}+u_a^{-1}u_b+u_a^{-1}u_b^{-1}).&
\end{flalign}

\subsection{Haar Measures}

The characters are very convenient objects.  One of their nice features is that they reduce tensor products of representations to ordinary products.  For example, $\chi_{SU(N),Adj}(u)=\chi_{SU(N),\,\fund{.4}}(u)\chi_{SU(N),\,\overline{\fund{.4}}}(u)-1$.  This makes them particularly useful when one wants to consider complicated compositions of representations.  The Haar measure provides the answer to the very specific but very well-motivated question, ``How many singlets are in a given composition?''
\be
\int_Hd\mu(z)\prod_{a=1}^k\chi_{H,\l_a}(z)=\,\,\mbox{number of singlets in}\,\,\bigotimes_{a=1}^{k}\l_a
\ee
The explicit forms of the measures for the groups that we need are
\begin{align}
\int_{SU(N)}f(z)d\mu(z)&=\frac1{N!}\int f(z)\Delta(z)\Delta(z^{-1})\prod_{a=1}^{N-1}\frac{dz_a}{2\pi iz_a}\bigg|_{\prod_{a=1}^Nz_a=1},\\
\int_{Sp(2N)}f(z)d\mu(z)&=\frac{(-1)^N}{2^NN!}\int f(z)\Delta(z+z^{-1})^2\prod_{i=1}^N\frac{dz_i}{z_i}(z_i-z_i^{-1})^2,
\end{align}
where
\be
\Delta(x)=\prod_{1\le a<b\le N}(x_a-x_b).
\ee

\section{On Total Ellipticity}\label{ellapp}

In this appendix, some details of the derivation of \eqref{ggg}-\eqref{g} are presented along with some comments on remaining confusions.  For the reader's convenience the quasi-periodicity conditions are reproduced:
\begin{align}\label{quasia}
\t(p^Au;p;q)_C&=(-u)^{-AC}q^{-AC(C-1)/2}p^{-C A(A-1)/2}\t(u;p;q)_C\\
\t(u;p;p^Bq)_C&=(-u)^{-BC(C-1)/2}q^{-BC(C-1)(2C-1)/6}p^{-BC(C-1)[B(2C-1)-3]/12}\t(u;p;q)_C\\
\t(p^Au;p;p^Bq)_C
&=(-u)^{-AC-BC(C-1)/2}q^{-AC(C-1)/2-BC(C-1)(2C-1)/6}\nonumber\\
&\times p^{-CA(A-1)/2-ABC(C-1)/2-BC(C-1)[B(2C-1)-3]/12}\t(u;p;q)_C
\end{align}
We will demand
\be
h_a(u_0,\dots,pu_b,\dots;p;q)=h_a(pu_0,\dots;p;p^Bq)=h_a(u;p;q),\qquad a\ne0
\ee
where
\be\label{q-certificatea}
h_a(u;p;q)=\prod_n\t(u^{A_n};p;q)^{\e_n}_{A_n^{(a)}},\qquad u^{A_n}=\prod_{a=0}^Mu_a^{A_n^{(a)}}
\ee

Using \eqref{quasia} we find the ellipticity conditions for the $u_b\ne u_0$ parameters,
\begin{align}
\prod_{n}\t(p^{A_n^{(b)}}u;p;q)_{A_n^{(a)}}^{\e_n}
=&(-1)^{\sum_{n}\e_nA_n^{(a)}A_n^{(b)}}\nonumber\\
&\times\prod_{C=0}^Mu_C^{-\sum_{n}\e_nA_n^{(a)}A_n^{(b)}A_n^{(C)}}\nonumber\\
&\times q^{-\sum_{n}\e_nA_n^{(b)}A_n^{(a)}(A_n^{(a)}-1)/2}\nonumber\\
&\times p^{-\sum_{n}\e_nA_n^{(a)}A_n^{(b)}(A_n^{(b)}-1)/2}\nonumber\\
&\times\prod_n\t(u^{A_n};p;q)_{A_n^{(a)}}^{\e_n}\nonumber\\
&=\prod_n\t(u^{A_n};p;q)_{A_n^{(a)}}^{\e_n}
\end{align}
Now combining the $u_0=(pq)^{1/B}$ factor with the $p$ and $q$ factors, we find
\begin{align}
&(-1)^{\sum_{n}\e_nA_n^{(a)}A_n^{(b)}}\nonumber\\
&\prod_{c=1}^Mu_k^{-\sum_{n}\e_nA_n^{(a)}A_n^{(b)}A_n^{(c)}}\nonumber\\
&\times q^{-\sum_{n}\e_nA_n^{(b)}A_n^{(a)}(A_n^{(a)}-1+\frac2BA_n^{(0)})/2}\nonumber\\
&\times p^{-\sum_{n}\e_nA_n^{(a)}A_n^{(b)}(A_n^{(b)}-1+\frac2BA_n^{(0)})/2}\nonumber\\
&=1
\end{align}
The vanishing of the exponents of the $u_c$ implies
\be\label{constraint1}
\sum_n\e_nA_n^{(a)}A_n^{(b)}A_n^{(c)}=0
\ee
This simplifies the exponents of $p$ and $q$, but leaves us with the condition
\be\label{constraint2}
\sum_n\e_nA_n^{(a)}A_n^{(b)}\left(2\frac{A_n^{(0)}}B-1\right)=0.
\ee
Recall that $A^{(0)}$ shows up in the factor $u_0^{A_n^{(0)}}=(pq)^{{A_n^{(0)}}/B}$, and when these factors arise in physical applications, they have the form $(pq)^{r/2}$, where $r$ is the $R$-charge of a chiral superfield.  This means that the quantity in parenthesis above gives the $R$-charge of the fermionic component of a chiral superfield,
\be\label{constraint3}
2\frac{A_n^{(0)}}{B}-1=R_{\psi_n},
\ee

Finally, fixing the sign implies the peculiar identity, 
\be\label{aaz}
\frac12\sum_n\e_nA_n^{(a)}A_n^{(b)}\in\mathbb Z.
\ee
We will come back to this.

Turning now to the conditions for ellipticity in $u_0$, we have
\begin{align}
\prod_n&\t(p^{A_n^{(0)}}u;p;p^Bq)_{A_n^{(a)}}^{\e_n}\nonumber\\
=&(-1)^{\sum_n\e_n[A_n^{(0)}A_n^{(a)}+BA_n^{(a)}(A_n^{(a)}-1)/2]}\nonumber\\
&\times\prod_{C=0}^nu_C^{-\sum_n\e_n[A_n^{(0)}A_n^{(a)}+BA_n^{(a)}(A_n^{(a)}-1)/2]A_n^{(C)}}\nonumber\\
&\times q^{-\sum_n\e_n[A_n^{(0)}A_n^{(a)}(A_n^{(a)}-1)/2+BA_n^{(a)}(A_n^{(a)}-1)(2A_n^{(a)}-1)/6]}\nonumber\\
&\times p^{-\sum_n\e_n[A_n^{(a)}A_n^{(0)}(A_n^{(0)}-1)/2+A_n^{(0)}BA_n^{(a)}(A_n^{(a)}-1)/2+BA_n^{(a)}(A_n^{(a)}-1)[B(2A_n^{(a)}-1)-3]/12]}\nonumber\\
&\times\prod_n\t(u^{A_n};p;q)_{A_n^{(a)}}^{\e_n}\nonumber\\
&=\prod_n\t(u^{A_n};p;q)_{A_n^{(a)}}^{\e_n}
\end{align}
Combining $u_0$ with the other $p$ and $q$ factors gives the following exponents (with some overall signs dropped).
\begin{align}
-1:&\sum_n\e_n\left[A_n^{(0)}A_n^{(a)}+\frac12BA_n^{(a)}(A_n^{(a)}-1)\right]\label{sign}\\
u_c:&\sum_n\e_n\left[A_n^{(0)}A_n^{(a)}+\frac12BA_n^{(a)}(A_n^{(a)}-1)\right]A_n^{(c)}\label{ufactor}\\
q:&\sum_n\e_n\bigg[\frac12A_n^{(0)}A_n^{(a)}(A_n^{(a)}-1)+\frac16BA_n^{(a)}(A_n^{(a)}-1)(2A_n^{(a)}-1)\nonumber\\
&\hspace{.5in}+[A_n^{(0)}A_n^{(a)}+\frac12BA_n^{(a)}(A_n^{(a)}-1)]\frac{A_n^{(0)}}{B}\bigg]\label{qfactor}\\
p:&\sum_n\e_n\bigg[\frac12A_n^{(a)}A_n^{(0)}(A_n^{(0)}-1)+\frac12A_n^{(0)}BA_n^{(a)}(A_n^{(a)}-1)\nonumber\\
&\hspace{.5in}+\frac1{12}BA_n^{(a)}(A_n^{(a)}-1)[B(2A_n^{(a)}-1)-3]+[A_n^{(0)}A_n^{(a)}+\frac12BA_n^{(a)}(A_n^{(a)}-1)]\frac{A_n^{(0)}}{B}\bigg]\\\nonumber
\end{align}
As before, $c\ne0$ and $C$ runs over all values including 0.

Applying our constraints, \eqref{constraint1} and \eqref{constraint2}, we find that the vanishing of the $u_c$ exponent \eqref{ufactor} gives nothing new.  The $q$ exponent \eqref{qfactor} gives
\be\label{constraint4}
\sum_n\e_nA_n^{(a)}(B^2-6BA_n^{(0)}+6A_n^{(0)}A_n^{(0)})=0.
\ee
Then using \eqref{constraint1}, \eqref{constraint2}, and \eqref{constraint4}, we find that the $p$ factor constraint simplifies to
\be\label{constraint5}
\sum_n\e_nA_n^{(a)}\left(2\frac{A_n^{(0)}}B-1\right)^2=0.
\ee
Feeding \eqref{constraint5} back into \eqref{constraint4}, we find that the remaining independent constraint is
\be\label{constraint6}
\sum_n\e_nA_n^{(a)}=0.
\ee
Finally, the sign \eqref{sign} implies
\be
\sum_n\frac12\e_n\left(A_n^{(a)}A_n^{(0)}+\frac B2A_n^{(a)}A_n^{(a)}\right)\in\mathbb Z.
\ee
This does not appear to be physical because it does not permit $B$ and $A_n^{(0)}$ to be rescaled by an arbitrary common factor.  If we aren't going to trust this sign, we shouldn't trust the other one \eqref{aaz}.  This doesn't appear to be consequential.  Overall signs don't have anything to do with whether one can perform an integral.

The more interesting question is, can the definition of total ellipticity be extended to include 't Hooft anomaly matching for $\Tr U(1)$ and $\Tr U(1)^3$ without spoiling the constraints \eqref{ggg}-\eqref{g}.  

\begin{landscape}
\section{$SU(N)$ Adjoint Deconfinement Module}\label{adjapp}
The following theory (above the double horizontal line), which confines to a single free $SU(N)$ adjoint (below the double line), was identified in \cite{Luty:1996cg}.  I prove in this appendix that the associated indices are equal.
\begin{center}
\begin{tabular}{c|c||c|c|c|c|c}
&\!\!$SU(\wt N)\!\!$&\!\!$SU(N)$\!\!&\!\!$SU(K)_1$\!\!&\!\!$SU(K)_2$\!\!&$U(1)$&$U(1)_R$\\\hline
$\varphi$&$\fund{.5}$&$\fund{.5}$&1&$1$&$e$&$r$\\\hline
$\wt \varphi$&$\overline{\fund{.5}}$&$\overline{\fund{.5}}$&1&1&$\tilde e$&$\tilde r$\\\hline
$Z_1$&$\fund{.5}$&1&$\fund{.5}$&1&$e_Z$&$r_Z$\\\hline
$\wt Z_1$&$\overline{\fund{.5}}$&1&1&$\overline{\fund{.5}}$&$-[ N(e + \tilde e)+K e_Z ]/K$&$[2\!-\!N(r+\tilde r)\!-\!Kr_Z]/K$\\\hline
$Z_2$&1&1&1&1&$-e-\tilde e$&$2-r-\tilde r$\\\hline
$Z_3$&1&$\overline{\fund{.5}}$&1&$\fund{.5}$&$[(N\!-\!K)e+N\tilde e+Ke_Z]/K$&\!\!$[2(K\!-\!1)+(N\!-\!K)r+N\tilde r+Kr_Z]/K$\!\!\\\hline
$\wt Z_3$&1&$\fund{.5}$&$\overline{\fund{.5}}$&1&$-\tilde e-e_Z$&$2 - \tilde r-r_Z$\\\hline
$Z_4$&1&1&$\overline{\fund{.5}}$&$\fund{.5}$&$N(e+\tilde e)/K$&$[2(K\!-\!1)+N(r+\tilde r)]/K$\\\hline
$Z_5$&1&$\fund{.5}$&1&1&$(1\!-\!N)e\!-\!Ke_Z$&$2+(1\!-\!N)r\!-\!Kr_Z $\\\hline
$\wt Z_5$&1&$\overline{\fund{.5}}$&1&1&$Ne+\tilde e+Ke_Z$&$Nr+\tilde r+Kr_Z $\\\hline
$Z_6$&1&1&$\fund{.5}$&1&$-Ne+(1\!-\!K)e_Z$&$2\!-\!Nr+(1\!-\!K)r_Z $\\\hline
$\wt Z_6$&1&1&1&$\overline{\fund{.5}}$&\!\!$[N(K\!-\!1)e\!-\!N\tilde e+K(K\!-\!1)e_Z]/K$\!\!&\!\!$[2+N(K\!-\!1)r\!-\!N\tilde r+K(K\!-\!1)r_Z]/K$\!\!\!\\\hline\hline
$\Phi$&$\cdot$&$Adj$&1&$1$&$e+\tilde e$&$r+\tilde r$\\
\end{tabular}
\end{center}
\be
\wt N=N+K-1
\ee
\be
W_{UV}=Z_2\varphi\wt\varphi+Z_3\varphi\wt Z_1+\wt Z_3\wt\varphi Z_1+Z_4Z_1\wt Z_1+Z_5\varphi^{N-1}Z_1^K+\wt Z_5\wt\varphi^{N-1}\wt Z_1^K+Z_6\varphi^NZ_1^{K-1}+\wt Z_6\wt\varphi^N\wt Z_1^{K-1}
\ee
\end{landscape}
{\vskip-.2in}
\be
U(1):y_1,\qquad SU(N):y_2,\qquad SU(K)_1:y_3,\qquad SU(K)_2:y_4
\ee
{\vskip-.1in}
With the above fugacity assignments the single field contributions are
\begin{flalign}
\qquad\I_\varphi&=\prod_{a=1}^N\prod_{b=1}^{\wt N}\Gamma\big((pq)^{r/2}y_1^ey_{2,a}z_b\big)&\\
\I_{\wt\varphi}&=\prod_{a=1}^N\prod_{b=1}^{\wt N}\Gamma\big((pq)^{\tilde r/2}y_1^{\tilde e}y_{2,a}^{-1}z_b^{-1}\big)&\\
\I_{Z_1}&=\prod_{a=1}^K\prod_{b=1}^{\wt N}\Gamma\big((pq)^{r_Z/2}y_1^{e_Z}y_{3,a}z_b\big)&\\
\I_{\wt Z_1}&=\prod_{a=1}^K\prod_{b=1}^{\wt N}\Gamma\big((pq)^{[2-N(r+\tilde r)-Kr_Z]/(2K)}y_1^{-[N(e+\tilde e)+Ke_Z]/K}y_{4,a}^{-1}z_b^{-1}\big)&\\
\I_{Z_2}&=\Gamma\big((pq)^{(2-r-\tilde r)/2}y_1^{-e-\tilde e}\big)&\\
\I_{Z_3}&=\prod_{a=1}^N\prod_{b=1}^K\Gamma\big((pq)^{[2(K-1)+(N-K)r+N\tilde r+Kr_Z]/(2K)}y_1^{[(N-K)e+N\tilde e+Ke_Z]/K}y_{2,a}^{-1}y_{4,b}\big)&\\
\I_{\wt Z_3}&=\prod_{a=1}^N\prod_{b=1}^K\Gamma\big((pq)^{(2-\tilde r-r_Z)/2}y_1^{-\tilde e-e_Z}y_{2,a}y_{3,b}^{-1}\big)&\\
\I_{Z_4}&=\prod_{a=1}^K\prod_{b=1}^K\Gamma\big((pq)^{[2(K-1)+N(r+\tilde r)]/(2K)}y_1^{N(e+\tilde e)/K}y_{3,a}^{-1}y_{4,b}\big)&\\
\I_{Z_5}&=\prod_{a=1}^N\Gamma\big((pq)^{[2+(1-N)r-Kr_Z]/2}y_1^{(1-N)e-Ke_Z}y_{2,a}\big)&\\
\I_{\wt Z_5}&=\prod_{a=1}^N\Gamma\big((pq)^{[Nr+\tilde r+Kr_Z]/2}y_1^{Ne+\tilde e+Ke_Z}y_{2,a}^{-1}\big)&\\
\I_{Z_6}&=\prod_{a=1}^K\Gamma\big((pq)^{[2-Nr+(1-K)r_Z]/2}y_1^{-Ne+(1-K)e_Z}y_{3,a}\big)&\\
\I_{\wt Z_6}&=\prod_{a=1}^K\Gamma\big((pq)^{[2+N(K-1)r-N\tilde r+K(K-1)r_Z]/(2K)}y_1^{[N(K-1)e-N\tilde e+K(K-1)e_Z]/K}y_{4,a}^{-1}\big)&\\
\I_{\Phi}&=\Gamma\big((pq)^{r+\tilde r}y_1^{e+\tilde e}\big)^{N-1}\prod_{1\le a<b\le N}\Gamma\big((pq)^{r+\tilde r}y_1^{e+\tilde e}y_{2,a}y_{2,b}^{-1},(pq)^{r+\tilde r}y_1^{e+\tilde e}y_{2,a}^{-1}y_{2,b}\big)
\end{flalign}
{\vskip-.1in}
I'll now show that the index equality of the $SU(N)$ adjoint deconfinement module,
\be
\I_{Z_2}\I_{Z_3}\I_{\wt Z_3}\I_{Z_4}\I_{Z_5}\I_{\wt Z_5}\I_{Z_6}\I_{\wt Z_6}\int[dz]_{SU(\wt N)}\I_\varphi\I_{\wt\varphi}\I_{Z_1}\I_{\wt Z_1}=\I_\Phi,
\ee 
follows from the basic $SU$ s-confinement integral identity.  We already used the $Sp$ s-confinement result in section \ref{antisec} and subsection \ref{cehtmag}.  The $SU$ version applies to the theory summarized in \eqref{suseiberg} when $N_f=N_c+1$.  The form of this integral is  
\be\label{cor4.2}
\int[dz]_{SU(N_c)}\prod_{a=1}^{N_c}\prod_{b=1}^{N_c+1}\Gamma(u_bz_a,\tilde u_bz_a^{-1})=\prod_{a=1}^{N_c+1}\Gamma(Uu_a^{-1},\wt U\tilde u_a^{-1})\prod_{b=1}^{N_c+1}\Gamma(u_a\tilde u_b),\qquad\prod_{a=1}^{N_c+1}u_a\tilde u_a=pq.
\ee
The definitions $U\equiv\prod_{a=1}^{N_c+1}u$ and $\wt U\equiv\prod_{a=1}^{N_c+1}\tilde u$ have been employed.
This identity was conjectured by Spiridonov \cite{spiriTwo} and proven by Rains (corollary 4.2 of \cite{2003math9252R}).  We start by re-writing the quark- and anti-quark-like fields,
\begin{align}
\I_\varphi\I_{Z_1}&=\prod_{a=1}^{\wt N}\prod_{b=1}^{\wt N+1}\Gamma(u_bz_a),\quad\boldsymbol{u}=\big((pq)^{r/2}y_1^e\boldsymbol{y_2},(pq)^{r_Z/2}y_1^{e_Z}\boldsymbol{y_3}\big)\\
\I_{\tilde\varphi}\I_{\wt Z_1}&=\prod_{a=1}^{\wt N}\prod_{b=1}^{\wt N+1}\Gamma(\tilde u_bz_a^{-1}),\quad\boldsymbol{\tilde u}=\big((pq)^{\tilde r/2}y_1^{\tilde e}\boldsymbol{y_2^{-1}},(pq)^{[2-N(r+\tilde r)-Kr_Z]/(2K)}y_1^{-[N(e+\tilde e)+Ke_Z]/K}\boldsymbol{y_4^{-1}}\big).
\end{align}
One can easily verify that $\prod_{a=1}^{\tilde N+1}u_a\tilde u_a=pq$.  We have reproduced the left side of \eqref{cor4.2}.  The remaining pieces can be similarly organized.
\be
\I_{Z_2}^{-1}\I_\Phi=\prod_{a=1}^N\prod_{b=1}^N\Gamma(\tilde u_au_b)
\ee
\be
\I_{Z_3}^{-1}=\prod_{a=N+1}^{K}\prod_{b=1}^{N}\Gamma(\tilde u_au_b)
\ee
\be
\I_{\wt Z_3}^{-1}=\prod_{a=1}^{N}\prod_{b=N+1}^{N+K}\Gamma(\tilde u_au_b)
\ee
\be
\I_{Z_4}^{-1}=\prod_{a=N+1}^{N+K}\prod_{b=N+1}^{N+K}\Gamma(\tilde u_au_b)
\ee
These four combine to give the double product on the right side of \eqref{cor4.2}.  The remaining pieces are
\be
\I_{Z_5}^{-1}=\prod_{a=1}^N\Gamma(Uu_a^{-1}),
\ee
\be
\I_{\wt Z_5}^{-1}=\prod_{a=1}^N\Gamma(\wt U\tilde u_a^{-1}),
\ee
\be
\I_{Z_6}^{-1}=\prod_{a=N+1}^{N+K}\Gamma(Uu_a^{-1}),
\ee
\be
\I_{\wt Z_6}^{-1}=\prod_{a=N+1}^{N+K}\Gamma(\wt U\tilde u_a^{-1}),
\ee
which combine to give the single product in \eqref{cor4.2}.  This shows that
\begin{align}
\int[dz]_{SU(\wt N)}\I_\varphi\I_{\wt\varphi}\I_{Z_1}\I_{\wt Z_1}\!&=\!\int[dz]_{SU(\wt N)}\prod_{a=1}^{\wt N}\prod_{b=1}^{\wt N+1}\Gamma(u_bz_a,\tilde u_bz_a^{-1})\nonumber\\
&=\prod_{a=1}^{\wt N+1}\!\Gamma(Uu_a^{-1},\wt U\tilde u_a^{-1})\!\prod_{b=1}^{\wt N+1}\!\Gamma(u_a\tilde u_b)=\I_{Z_2}^{-1}\I_{Z_3}^{-1}\I_{\wt Z_3}^{-1}\I_{Z_4}^{-1}\I_{Z_5}^{-1}\I_{\wt Z_5}^{-1}\I_{Z_6}^{-1}\I_{\wt Z_6}^{-1}\I_{\Phi}.
\end{align}

\bibliographystyle{distler.bst}
\bibliography{torrobabib}

\providecommand{\href}[2]{#2}\begingroup\raggedright\begin{thebibliography}{10}

\bibitem{Wess:1973kz}
J.~Wess and B.~Zumino, ``{A Lagrangian Model Invariant Under Supergauge
  Transformations},''
\href{http://dx.doi.org/10.1016/0370-2693(74)90578-4}{{\em Phys.Lett.}
  {\bfseries B49} (1974) 52}.

\bibitem{Wess:1974jb}
J.~Wess and B.~Zumino, ``{Supergauge Invariant Extension of Quantum
  Electrodynamics},''
\href{http://dx.doi.org/10.1016/0550-3213(74)90112-6}{{\em Nucl.Phys.}
  {\bfseries B78} (1974) 1}.

\bibitem{Iliopoulos:1974zv}
J.~Iliopoulos and B.~Zumino, ``{Broken Supergauge Symmetry and
  Renormalization},''
\href{http://dx.doi.org/10.1016/0550-3213(74)90388-5}{{\em Nucl.Phys.}
  {\bfseries B76} (1974) 310}.

\bibitem{Grisaru:1979wc}
M.~T. Grisaru, W.~Siegel, and M.~Rocek, ``{Improved Methods for Supergraphs},''
\href{http://dx.doi.org/10.1016/0550-3213(79)90344-4}{{\em Nucl.Phys.}
  {\bfseries B159} (1979) 429}.

\bibitem{Hanany:1996ie}
A.~Hanany and E.~Witten, ``{Type IIB superstrings, BPS monopoles, and
  three-dimensional gauge dynamics},''
  \href{http://dx.doi.org/10.1016/S0550-3213(97)00157-0}{{\em Nucl.Phys.}
  {\bfseries B492} (1997) 152--190},
\href{http://arxiv.org/abs/hep-th/9611230}{{\ttfamily arXiv:hep-th/9611230
  [hep-th]}}.

\bibitem{Elitzur:1997fh}
S.~Elitzur, A.~Giveon, and D.~Kutasov, ``{Branes and N=1 duality in string
  theory},'' \href{http://dx.doi.org/10.1016/S0370-2693(97)00375-4}{{\em
  Phys.Lett.} {\bfseries B400} (1997) 269--274},
\href{http://arxiv.org/abs/hep-th/9702014}{{\ttfamily arXiv:hep-th/9702014
  [hep-th]}}.

\bibitem{Schmaltz:1997sq}
M.~Schmaltz and R.~Sundrum, ``{N=1 field theory duality from M theory},''
  \href{http://dx.doi.org/10.1103/PhysRevD.57.6455}{{\em Phys.Rev.} {\bfseries
  D57} (1998) 6455--6463},
\href{http://arxiv.org/abs/hep-th/9708015}{{\ttfamily arXiv:hep-th/9708015
  [hep-th]}}.

\bibitem{Csaki:1997qe}
C.~Csaki and W.~Skiba, ``{Duality in Sp and SO gauge groups from M theory},''
  \href{http://dx.doi.org/10.1016/S0370-2693(97)01222-7}{{\em Phys. Lett.}
  {\bfseries B415} (1997) 31--38},
\href{http://arxiv.org/abs/hep-th/9708082}{{\ttfamily arXiv:hep-th/9708082}}.

\bibitem{Pouliot:1998yv}
P.~Pouliot, ``{Molien function for duality},'' {\em JHEP} {\bfseries 9901}
  (1999) 021,
\href{http://arxiv.org/abs/hep-th/9812015}{{\ttfamily arXiv:hep-th/9812015
  [hep-th]}}.

\bibitem{Benvenuti:2006qr}
S.~Benvenuti, B.~Feng, A.~Hanany, and Y.-H. He, ``{Counting BPS Operators in
  Gauge Theories: Quivers, Syzygies and Plethystics},''
  \href{http://dx.doi.org/10.1088/1126-6708/2007/11/050}{{\em JHEP} {\bfseries
  0711} (2007) 050},
\href{http://arxiv.org/abs/hep-th/0608050}{{\ttfamily arXiv:hep-th/0608050
  [hep-th]}}.

\bibitem{Feng:2007ur}
B.~Feng, A.~Hanany, and Y.-H. He, ``{Counting gauge invariants: The Plethystic
  program},'' \href{http://dx.doi.org/10.1088/1126-6708/2007/03/090}{{\em JHEP}
  {\bfseries 0703} (2007) 090},
\href{http://arxiv.org/abs/hep-th/0701063}{{\ttfamily arXiv:hep-th/0701063
  [hep-th]}}.

\bibitem{Chen:2011wn}
Y.~Chen and N.~Mekareeya, ``{The Hilbert series of U/SU SQCD and Toeplitz
  Determinants},''
  \href{http://dx.doi.org/10.1016/j.nuclphysb.2011.05.012}{{\em Nucl.Phys.}
  {\bfseries B850} (2011) 553--593},
\href{http://arxiv.org/abs/1104.2045}{{\ttfamily arXiv:1104.2045 [hep-th]}}.

\bibitem{Witten:1982df}
E.~Witten, ``{Constraints on Supersymmetry Breaking},''
\href{http://dx.doi.org/10.1016/0550-3213(82)90071-2}{{\em Nucl.Phys.}
  {\bfseries B202} (1982) 253}.

\bibitem{Sen:1985ph}
D.~Sen, ``{SUPERSYMMETRY IN THE SPACE-TIME R x S**3},''
\href{http://dx.doi.org/10.1016/0550-3213(87)90033-2}{{\em Nucl.Phys.}
  {\bfseries B284} (1987) 201}.

\bibitem{Sen:1989bg}
D.~Sen, ``{EXTENDED SUPERSYMMETRY IN THE SPACE-TIME R x S**3},''
\href{http://dx.doi.org/10.1103/PhysRevD.41.667}{{\em Phys.Rev.} {\bfseries
  D41} (1990) 667}.

\bibitem{Festuccia:2011ws}
G.~Festuccia and N.~Seiberg, ``{Rigid Supersymmetric Theories in Curved
  Superspace},'' \href{http://dx.doi.org/10.1007/JHEP06(2011)114}{{\em JHEP}
  {\bfseries 1106} (2011) 114},
\href{http://arxiv.org/abs/1105.0689}{{\ttfamily arXiv:1105.0689 [hep-th]}}.

\bibitem{Romelsberger:2005eg}
C.~Romelsberger, ``{Counting chiral primaries in N = 1, d=4 superconformal
  field theories},''
  \href{http://dx.doi.org/10.1016/j.nuclphysb.2006.03.037}{{\em Nucl.Phys.}
  {\bfseries B747} (2006) 329--353},
  \href{http://arxiv.org/abs/hep-th/0510060}{{\ttfamily arXiv:hep-th/0510060
  [hep-th]}}.

\bibitem{Kinney:2005ej}
J.~Kinney, J.~M. Maldacena, S.~Minwalla, and S.~Raju, ``{An Index for 4
  dimensional super conformal theories},''
  \href{http://dx.doi.org/10.1007/s00220-007-0258-7}{{\em Commun.Math.Phys.}
  {\bfseries 275} (2007) 209--254},
  \href{http://arxiv.org/abs/hep-th/0510251}{{\ttfamily arXiv:hep-th/0510251
  [hep-th]}}.

\bibitem{Romelsberger:2007ec}
C.~Romelsberger, ``{Calculating the Superconformal Index and Seiberg
  Duality},'' \href{http://arxiv.org/abs/0707.3702}{{\ttfamily arXiv:0707.3702
  [hep-th]}}.

\bibitem{Zwiebel:2011wa}
B.~I. Zwiebel, ``{Charging the Superconformal Index},''
\href{http://arxiv.org/abs/1111.1773}{{\ttfamily arXiv:1111.1773 [hep-th]}}.

\bibitem{Dolan:2008qi}
F.~Dolan and H.~Osborn, ``{Applications of the Superconformal Index for
  Protected Operators and q-Hypergeometric Identities to N=1 Dual Theories},''
  \href{http://dx.doi.org/10.1016/j.nuclphysb.2009.01.028}{{\em Nucl.Phys.}
  {\bfseries B818} (2009) 137--178},
  \href{http://arxiv.org/abs/0801.4947}{{\ttfamily arXiv:0801.4947 [hep-th]}}.

\bibitem{anon}
Anonymous {\em \!\!, private communication\!} .

\bibitem{Gadde:2010en}
A.~Gadde, L.~Rastelli, S.~S. Razamat, and W.~Yan, ``{On the Superconformal
  Index of N=1 IR Fixed Points: A Holographic Check},''
  \href{http://dx.doi.org/10.1007/JHEP03(2011)041}{{\em JHEP} {\bfseries 1103}
  (2011) 041},
\href{http://arxiv.org/abs/1011.5278}{{\ttfamily arXiv:1011.5278 [hep-th]}}.

\bibitem{Spiridonov:2008zr}
V.~Spiridonov and G.~Vartanov, ``{Superconformal indices for N = 1 theories
  with multiple duals},''
  \href{http://dx.doi.org/10.1016/j.nuclphysb.2009.08.022}{{\em Nucl.Phys.}
  {\bfseries B824} (2010) 192--216},
  \href{http://arxiv.org/abs/0811.1909}{{\ttfamily arXiv:0811.1909 [hep-th]}}.

\bibitem{Spiridonov:2009za}
V.~Spiridonov and G.~Vartanov, ``{Elliptic Hypergeometry of Supersymmetric
  Dualities},'' \href{http://dx.doi.org/10.1007/s00220-011-1218-9}{{\em
  Commun.Math.Phys.} {\bfseries 304} (2011) 797--874},
  \href{http://arxiv.org/abs/0910.5944}{{\ttfamily arXiv:0910.5944 [hep-th]}}.

\bibitem{Spiridonov:2011hf}
V.~Spiridonov and G.~Vartanov, ``{Elliptic hypergeometry of supersymmetric
  dualities II. Orthogonal groups, knots, and vortices},''
  \href{http://arxiv.org/abs/1107.5788}{{\ttfamily arXiv:1107.5788 [hep-th]}}.

\bibitem{Seiberg:1994pq}
N.~Seiberg, ``{Electric - magnetic duality in supersymmetric nonAbelian gauge
  theories},'' \href{http://dx.doi.org/10.1016/0550-3213(94)00023-8}{{\em
  Nucl.Phys.} {\bfseries B435} (1995) 129--146},
\href{http://arxiv.org/abs/hep-th/9411149}{{\ttfamily arXiv:hep-th/9411149
  [hep-th]}}.

\bibitem{Intriligator:1995ne}
K.~A. Intriligator and P.~Pouliot, ``{Exact superpotentials, quantum vacua and
  duality in supersymmetric SP(N(c)) gauge theories},''
  \href{http://dx.doi.org/10.1016/0370-2693(95)00618-U}{{\em Phys.Lett.}
  {\bfseries B353} (1995) 471--476},
\href{http://arxiv.org/abs/hep-th/9505006}{{\ttfamily arXiv:hep-th/9505006
  [hep-th]}}.

\bibitem{Kutasov:1995ve}
D.~Kutasov, ``{A Comment on duality in N=1 supersymmetric nonAbelian gauge
  theories},'' \href{http://dx.doi.org/10.1016/0370-2693(95)00392-X}{{\em
  Phys.Lett.} {\bfseries B351} (1995) 230--234},
\href{http://arxiv.org/abs/hep-th/9503086}{{\ttfamily arXiv:hep-th/9503086
  [hep-th]}}.

\bibitem{Kutasov:1995np}
D.~Kutasov and A.~Schwimmer, ``{On duality in supersymmetric Yang-Mills
  theory},'' \href{http://dx.doi.org/10.1016/0370-2693(95)00676-C}{{\em
  Phys.Lett.} {\bfseries B354} (1995) 315--321},
\href{http://arxiv.org/abs/hep-th/9505004}{{\ttfamily arXiv:hep-th/9505004
  [hep-th]}}.

\bibitem{Kutasov:1995ss}
D.~Kutasov, A.~Schwimmer, and N.~Seiberg, ``{Chiral rings, singularity theory
  and electric - magnetic duality},''
  \href{http://dx.doi.org/10.1016/0550-3213(95)00599-4}{{\em Nucl.Phys.}
  {\bfseries B459} (1996) 455--496},
\href{http://arxiv.org/abs/hep-th/9510222}{{\ttfamily arXiv:hep-th/9510222
  [hep-th]}}.

\bibitem{spiriOne}
V.~Spiridonov, ``{On the elliptic beta function},'' {\em Russ. Math. Surveys}
  {\bfseries 56} (2001) 185--186.

\bibitem{spiriTwo}
V.~P. {Spiridonov}, ``{Theta hypergeometric integrals},'' {\em Algebra i
  Analiz} {\bfseries 15} (2003) 161--215,
  \href{http://arxiv.org/abs/arXiv:math/0303205}{{\ttfamily
  arXiv:math/0303205}}.

\bibitem{2003math9252R}
E.~M. {Rains}, ``{Transformations of elliptic hypergometric integrals},'' {\em
  ArXiv Mathematics e-prints} (Sept., 2003) ,
  \href{http://arxiv.org/abs/arXiv:math/0309252}{{\ttfamily
  arXiv:math/0309252}}.

\bibitem{2010arXiv1003.4491S}
V.~P. {Spiridonov}, ``{Elliptic hypergeometric terms},'' {\em ArXiv e-prints}
  (Mar., 2010) , \href{http://arxiv.org/abs/1003.4491}{{\ttfamily
  arXiv:1003.4491 [math.CA]}}.

\bibitem{'tHooft:1980xb}
e.~'t~Hooft, Gerard, e.~Itzykson, C., e.~Jaffe, A., e.~Lehmann, H., e.~Mitter,
  P.K., {\em et al.}, ``{Recent Developments in Gauge Theories. Proceedings,
  Nato Advanced Study Institute, Cargese, France, August 26 - September 8,
  1979},''
{\em NATO Adv.Study Inst.Ser.B Phys.} {\bfseries 59} (1980) 1--438.

\bibitem{Vartanov:2010xj}
G.~Vartanov, ``{On the ISS model of dynamical SUSY breaking},''
  \href{http://dx.doi.org/10.1016/j.physletb.2010.12.040}{{\em Phys.Lett.}
  {\bfseries B696} (2011) 288--290},
  \href{http://arxiv.org/abs/1009.2153}{{\ttfamily arXiv:1009.2153 [hep-th]}}.

\bibitem{Khmelnitsky:2009vc}
A.~Khmelnitsky, ``{Interpreting multiple dualities conjectured from
  superconformal index identities},''
  \href{http://dx.doi.org/10.1007/JHEP03(2010)065}{{\em JHEP} {\bfseries 1003}
  (2010) 065},
\href{http://arxiv.org/abs/0912.4523}{{\ttfamily arXiv:0912.4523 [hep-th]}}.

\bibitem{Spiridonov:2010hh}
V.~Spiridonov and G.~Vartanov, ``{Supersymmetric dualities beyond the conformal
  window},'' \href{http://dx.doi.org/10.1103/PhysRevLett.105.061603}{{\em
  Phys.Rev.Lett.} {\bfseries 105} (2010) 061603},
  \href{http://arxiv.org/abs/1003.6109}{{\ttfamily arXiv:1003.6109 [hep-th]}}.

\bibitem{Berkooz:1995km}
M.~Berkooz, ``{The Dual of supersymmetric SU(2k) with an antisymmetric tensor
  and composite dualities},''
  \href{http://dx.doi.org/10.1016/0550-3213(95)00400-M}{{\em Nucl.Phys.}
  {\bfseries B452} (1995) 513--525},
\href{http://arxiv.org/abs/hep-th/9505067}{{\ttfamily arXiv:hep-th/9505067
  [hep-th]}}.

\bibitem{Intriligator:1995id}
K.~A. Intriligator and N.~Seiberg, ``{Duality, monopoles, dyons, confinement
  and oblique confinement in supersymmetric SO(N(c)) gauge theories},''
  \href{http://dx.doi.org/10.1016/0550-3213(95)00159-P}{{\em Nucl.Phys.}
  {\bfseries B444} (1995) 125--160},
\href{http://arxiv.org/abs/hep-th/9503179}{{\ttfamily arXiv:hep-th/9503179
  [hep-th]}}.

\bibitem{Luty:1996cg}
M.~A. Luty, M.~Schmaltz, and J.~Terning, ``{A Sequence of duals for Sp(2N)
  supersymmetric gauge theories with adjoint matter},''
  \href{http://dx.doi.org/10.1103/PhysRevD.54.7815}{{\em Phys.Rev.} {\bfseries
  D54} (1996) 7815--7824},
\href{http://arxiv.org/abs/hep-th/9603034}{{\ttfamily arXiv:hep-th/9603034
  [hep-th]}}.

\bibitem{2004TMP139536S}
V.~P. {Spiridonov}, ``{A Bailey Tree for Integrals},''
  \href{http://dx.doi.org/10.1023/B:TAMP.0000022745.45082.18}{{\em Theoretical
  and Mathematical Physics} {\bfseries 139} (Apr., 2004) 536--541},
  \href{http://arxiv.org/abs/arXiv:math/0312502}{{\ttfamily
  arXiv:math/0312502}}.

\bibitem{Craig:2011tx}
N.~Craig, R.~Essig, A.~Hook, and G.~Torroba, ``{New dynamics and dualities in
  supersymmetric chiral gauge theories},''
  \href{http://dx.doi.org/10.1007/JHEP09(2011)046}{{\em JHEP} {\bfseries 1109}
  (2011) 046},
\href{http://arxiv.org/abs/1106.5051}{{\ttfamily arXiv:1106.5051 [hep-th]}}.

\bibitem{Craig:2011wj}
N.~Craig, R.~Essig, A.~Hook, and G.~Torroba, ``{Phases of N=1 supersymmetric
  chiral gauge theories},''
\href{http://arxiv.org/abs/1110.5905}{{\ttfamily arXiv:1110.5905 [hep-th]}}.

\bibitem{Intriligator:2003jj}
K.~A. Intriligator and B.~Wecht, ``{The Exact superconformal R symmetry
  maximizes a},'' \href{http://dx.doi.org/10.1016/S0550-3213(03)00459-0}{{\em
  Nucl.Phys.} {\bfseries B667} (2003) 183--200},
\href{http://arxiv.org/abs/hep-th/0304128}{{\ttfamily arXiv:hep-th/0304128
  [hep-th]}}.

\bibitem{1999math7061F}
G.~{Felder} and A.~{Varchenko}, ``{The elliptic gamma function and SL(3,Z) x
  Z\^{}3},'' {\em ArXiv Mathematics e-prints} (July, 1999) ,
  \href{http://arxiv.org/abs/arXiv:math/9907061}{{\ttfamily
  arXiv:math/9907061}}.

\bibitem{Intriligator:1995au}
K.~A. Intriligator and N.~Seiberg, ``{Lectures on supersymmetric gauge theories
  and electric - magnetic duality},'' {\em Nucl.Phys.Proc.Suppl.} {\bfseries
  45BC} (1996) 1--28,
\href{http://arxiv.org/abs/hep-th/9509066}{{\ttfamily arXiv:hep-th/9509066
  [hep-th]}}.

\end{thebibliography}\endgroup
\end{document}